\documentclass[letterpaper,twocolumn,10pt]{article}
\usepackage{usenix2019_v3}
\usepackage{amsfonts}

\begin{document}
%-------------------------------------------------------------------------------

% Don't want date printed.
\date{}

% NOTE: acmart-specific commands removed for USENIX:
%   \settopmatter, \setcopyright, \acmYear, \acmDOI, \acmConference, \acmISBN,
%   the \ifpreprint copyright toggle, CCSXML concepts, and \keywords.
%   USENIX does not use ACM rights metadata, CCS concepts, or a keywords block.

\title{\Large \bf LLM Ghostbusters: Surgical Package Hallucination Suppression via Adaptive Unlearning}

% USENIX Security submissions are typically anonymized. Comment the author
% block out for submission and restore it for the camera-ready.
\author{
{\rm Joseph Spracklen\textsuperscript{$\dagger$}}\\
University of Texas at San Antonio
\and
{\rm Pedram Aghazadeh\textsuperscript{$\dagger$}}\\
University of California San Diego
\and
{\rm Farinaz Koushanfar}\\
University of California San Diego
\and
{\rm Murtuza Jadliwala}\\
University of Texas at San Antonio
} % end author
% \author{
% {\rm Anonymous Author(s)}\\
% Anonymous Institution(s)
% } % end author

\maketitle\begingroup

\renewcommand{\thefootnote}{\fnsymbol{footnote}}
\footnotetext[1]{Equal contribution.}
\endgroup
%-------------------------------------------------------------------------------
\begin{abstract}

Hallucinations remain an unsolved problem for LLMs, and \emph{package hallucinations} are a particularly dangerous instance of this phenomenon. Package hallucinations occur during code generation when a model fabricates non-existent software packages, recommending imports and installation commands for fictional libraries. This creates a critical supply-chain vulnerability; an attacker can proactively register such packages on public registries with malicious payloads that are subsequently installed and executed by developers or autonomous agents. These hallucinations enable a class of package confusion attack known as \emph{slopsquatting}. To address this issue, we present \textbf{Adaptive Unlearning} (AU), a post-deployment framework that surgically suppresses package hallucinations while preserving general model utility. AU introduces a hybrid token-level objective that simultaneously reinforces valid outputs and suppresses hallucinated ones. Combined with an adaptive discovery loop that continuously surfaces new hallucination-inducing contexts without human supervision, AU enables generalization to unseen prompts and hallucinations. We demonstrate that AU reduces package hallucination rates by \textbf{88\%}, while maintaining performance on standard coding benchmarks. Our analysis shows that distributional changes are concentrated on package-related generations, leaving general coding behavior largely unaffected and confirming that AU's effect is isolated to the targeted distribution. AU relies entirely on model-generated data and requires no human annotation, representing a post-deployment hallucination mitigation framework.
\end{abstract}

%-------------------------------------------------------------------------------
\section{Introduction}
\label{sec:intro}

Large language models (LLMs) have moved from research artifacts into production infrastructure, with billions of daily interactions now routed through systems that synthesize information, answer questions, and generate code on behalf of users and downstream applications. This trajectory has made LLMs a load-bearing component of software supply chains, developer workflows, and enterprise decision systems and, in turn, has made their failure modes security-relevant. Most notable among these failures are \emph{hallucinations}, fluent and plausible outputs that are nonetheless factually incorrect, unsupported by the input, or fabricated.

% Large Language Models (LLMs) have rapidly transformed from interesting research into critical infrastructure powering billions of interactions daily. Their ability to generate human-quality text, synthesize complex information, and assist with sophisticated reasoning tasks has made them indispensable across industries. Yet, this success has created pressing challenges: as models are deployed at scale in high-stakes applications, from medical diagnosis to financial analysis to code generation, their failures become increasingly consequential. These failures, called hallucinations, are instances where models generate plausible sounding but factually incorrect, or even entirely fabricated, information. Hallucinations are an inherent problem with LLMs due to the stochastic nature of these systems and represent both a persistent and dangerous flaw.

In some cases, hallucinations have been shown to be exploitable vulnerabilities rather than merely quality defects. A prime example is a use case specific to code generation called package hallucination. When an LLM fabricates a package name that does not exist in the respective package repository, the hallucinated identifier becomes an attack target: an adversary can register the fabricated name with a malicious payload and wait for downstream users or autonomous coding agents to download and install it. This is the basis of \emph{slopsquatting} and \emph{package confusion} attacks against LLM-generated code~\cite{spracklen2025}. Any model output that names an external resource (packages, URLs, API endpoints, registry identifiers) creates an analogous surface whenever the hallucination is predictable or reproducible. In this setting, hallucination rate is a measure of attack-surface volume.

Will this surface shrink to zero as models improve? The evidence says no. Scaling model size and data, long the dominant recipe for capability gains~\cite{kaplan2020scaling, hoffmann2022training}, is running into both practical limits~\cite{villalobos2024runout}, and it does not appear to eliminate hallucinations~\cite{zheng2026llmsinventrustcrates}. TruthfulQA~\cite{lin-etal-2022-truthfulqa} documents cases where larger models become \emph{less} truthful, and recent theoretical work argues that hallucinations are not implementation bugs but artifacts of training a model to maximize likelihood rather than acknowledge uncertainty~\cite{kalai2025hallucinate}. Post-training refinements, such as RLHF~\cite{ouyang2022training}, RLAIF~\cite{bai2022constitutional}, GRPO~\cite{shao2024deepseekmathpushinglimitsmathematical}, are likewise insufficient. Recent analyses indicate that RL-based post-training predominantly reshapes output preferences and sampling behavior rather than altering the pretrained knowledge representations that generate hallucinations in the first place~\cite{yue2025doesreinforcementlearningreally, zhang2024negative}. 

Package hallucinations have emerged as a distinct and active research field and serve as an unofficial benchmark for hallucination progress due to their clear relevance as an active security threat and ease of detection. Since the first large scale characterization of the phenomenon~\cite{spracklen2025}, a growing body of dedicated work has extended the measurements to new metrics~\cite{krishna2025phantoms}, additional ecosystems~\cite{zheng2026rust}, and to additional prompt variations~\cite{twist2025library}. A parallel line has produced countermeasures spanning adversarial fuzzing~\cite{zhao2025hfuzzertestinglargelanguage}, decoding-time constraints~\cite{liu2026packmonitorenablingzeropackage}, localized model editing~\cite{liu2026mitigatingpackagehallucinationslarge}, and calibrated detection~\cite{hillah2026bayesiancalibrateddetectionhallucinatedpackage}. Crucially, this problem has not been overtaken by model scaling; recent research on the latest frontier models reports hallucination rates have not improved since the original study~\cite{churilov2026range} and identifies a large common set of hallucinated package names that appear in every evaluated frontier model. These latest results confirm that package hallucinations are a persistent and exploitable phenomenon that demands intervention beyond waiting for models to improve organically.

Comprehensive surveys~\cite{huang2024survey, zhang2023sirens} and measurement studies~\cite{min-etal-2023-factscore, li-etal-2023-halueval} confirm the pattern empirically: hallucinations persist across tasks, domains, and model generations, at rates high enough that frontier vendors now treat hallucination reduction as a primary release-gating objective. OpenAI's GPT-5 announcement centers a 45--80\% improvement on factuality benchmarks~\cite{openai2025gpt5}; Anthropic's Claude~4 release~\cite{anthropic2025claude4} and Google's Gemini~3 release~\cite{googledeepmind2025gemini3} each foreground hallucination behavior as a primary deployment concern, with Gemini~3 exhibiting an 88\% hallucination rate on a subset of queries where the model answers confidently rather than deferring.

Taken together, these observations motivate a different class of intervention. If hallucinations cannot be fully eliminated during pre-training and are only partially masked by post-training alignment, then reducing their security impact requires \emph{post-deployment refinement}. Mechanisms that precisely suppress specific unwanted behaviors in an already-trained model, without retraining from scratch while avoiding collateral damage to capability. This is the problem we address in this work.

\subsection{The Post-Deployment Refinement Gap}
\label{sec:gap}

An ideal post-deployment refinement mechanism would satisfy four properties simultaneously:

\begin{enumerate}
    \item \textbf{Precision.} Remove the targeted hallucination behavior with negligible degradation on unrelated capabilities.
    \item \textbf{Generalization.} Suppress hallucinations across prompt variations, rather than memorizing fixes for specific prompt--output pairs that an adversary or user could trivially rephrase.
    \item \textbf{Data efficiency.} Operate on synthetic data generated by the model itself, without requiring human-annotated hallucination corpora.
    \item \textbf{Scalability.} Support continuous refinement as new hallucinations are reported in production, on timescales compatible with release cycles
\end{enumerate}

Existing approaches fall short of these requirements. Privacy-focused machine unlearning methods~\cite{bourtoule2021machine, cao2015towards, guo2020certified} target removal of specific training examples to satisfy regulations like GDPR, but hallucinations are not discrete facts to be forgotten, they are rather emergent behaviors distributed throughout the model. Knowledge editing techniques like ROME~\cite{meng2022locating} and MEMIT~\cite{meng2023memit} can update specific factual associations, but they operate on the premise of \emph{replacing} rather than \emph{suppressing unwanted generation patterns}. Retrieval-augmented generation~\cite{lewis2020retrieval} and inference-time verification methods~\cite{manakul2023selfcheckgpt} add computational overhead and architectural complexity while failing to address the root cause in the model's learned representations. Constitutional AI and self-consistency approaches rely on a model's ability to critique itself, a capability that fails when models hallucinate confidently~\cite{huang2024selfcorrect,zhang2023sac3}.

Recent work on Partial Model Collapse (PMC)~\cite{scholten2025pmc} offers a promising avenue in Machine Unlearning. PMC recognizes that the same collapse dynamics that threaten recursive training regimes can be deliberately induced in a \emph{controlled} fashion to remove targeted content while preserving overall utility, turning an uncontrolled failure mode into a precise unlearning mechanism. However, PMC is designed for traditional machine-unlearning scenarios in which the target set is finite, well-defined, and known in advance (e.g., a specific set of training documents to forget due to privacy concerns). Hallucination reduction violates all three assumptions: the space of hallucination-inducing prompts is open-ended, the set of hallucinated outputs a model will produce is unknown until elicited, and new failure modes surface continuously as deployment contexts shift. Closing the gap to a deployable post-deployment refinement mechanism therefore requires two additional components PMC does not provide: (i) a mechanism to \emph{discover} hallucination-inducing prompts and (ii) a training scheme that generalizes suppression to prompt variations the method has not yet seen.

\begin{figure*}[!t]
  \centering
  \includegraphics[width=0.95\textwidth]{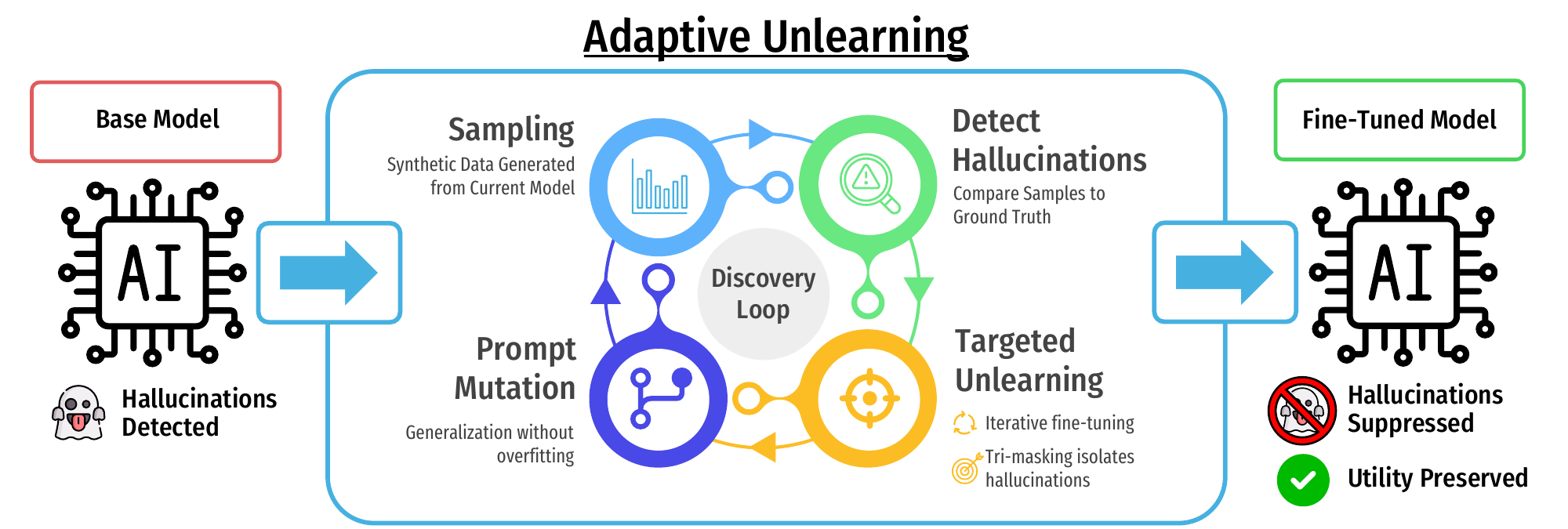}
  \caption{\textbf{Adaptive Unlearning pipeline.} The four-stage discovery loop, Sampling, Detection, Targeted Unlearning, and Prompt Mutation, iteratively suppresses hallucinated package tokens while preserving general coding utility. The full method is detailed in \S\ref{sec:method}.}
  \label{fig:pipeline}
\end{figure*}

\subsection{Our Approach.}

We address this gap with \textbf{Adaptive Unlearning (AU)}\footnote{Anonymized code available at : \url{https://anonymous.4open.science/r/Adaptive-Unlearning-952E}}, a closed-loop post-deployment refinement framework that directly supplies the two ingredients other methods such as PMC lack. First, an adaptive prompt-mutation loop that continuously elicits new hallucination-inducing contexts from the model itself, rather than assuming a fixed target set. Second, a token-level hybrid objective that reinforces valid outputs while suppressing hallucinated ones within the same sequence, generalizing suppression to prompt variations the method has not seen. AU extends the Model Collapse and NPO frameworks~\cite{scholten2025pmc, zhang2024negative} to the open-ended prompt space inherent in hallucination reduction, operates entirely on synthetic self-generated data, and requires no human annotated corpora. In our experiments, AU reduces hallucination rates by over \textbf{88\%} on unseen prompt variations while preserving utility on established code-generation benchmarks, as detailed in \S\ref{sec:method} and as shown in Fig.~\ref{fig:pipeline}.

\subsection{Contributions}

Our work makes the following contributions:

\begin{enumerate}

    \item \textbf{Adaptive Unlearning:} We introduce AU, a novel closed-loop post-deployment framework that surgically suppresses package hallucinations without full retraining, not requiring human annotation, and without a pre-specified forget set. AU is the first unlearning method to couple an adaptive hallucination-discovery loop with a hybrid token-level objective, enabling suppression that generalizes to unseen prompts rather than memorizing fixes for known ones. In package hallucination tests, a setting with direct security consequences via slopsquatting, AU reduces hallucination rates by 88\% while preserving coding benchmark performance, outperforming all evaluated baselines on the joint hallucination-utility tradeoff.

    \item \textbf{Tri-mask token routing:} AU uses a novel three-valued gradient-routing mechanism that partitions the tokens into three sets: valid-content tokens, hallucinated tokens, and context tokens. We use distinct loss terms for each. Unlike the binary masking used in prior unlearning work~\cite{liu2020learn, deng2025forgetitall}, tri-masking enables surgical edits within a single sequence without affecting adjacent tokens. This fine-grained routing is what allows the CE and NPO objectives run on separate token populations without interfering.

    \item \textbf{Adaptive Prompt Mutation:} We introduce a closed-loop mutation strategy that continuously elicits hallucination-inducing contexts from the model itself, retiring exhausted prompts in favor of semantically related variants. This extends existing unlearning methods from finite, pre-specified forget sets to the open-ended prompt space inherent in real-world settings, enabling suppression of hallucination patterns rather than memorization of prompt-output pairs.

    \item \textbf{Nested training for stable unlearning.} We propose a nested-loop training structure with outer resampling epochs and inner unlearning epochs that caches synthetic samples for multiple inner steps before regeneration. This directly addresses the instability of prior approaches that resample every step and therefore never consolidate knowledge. To the best of our knowledge, this nested cache-refresh scheme has not previously been applied to collapse-based unlearning, and our ablations show it is crucial for stable hallucination reduction.

    \item \textbf{Systematic joint evaluation.} We provide, to our knowledge, the first empirical comparison of unlearning methods for any type of hallucination reduction, evaluated jointly across hallucination rate, distributional drift, and coding utility. This joint view is necessary: methods that appear effective on any single axis often fail on another, and the tradeoff is only visible when all three are reported together.
\end{enumerate}
% \claude{Reviewer B calls the introduction ``alarmist'' and notes the cited
% hallucination-rate studies predate current models -- caveat that rates have
% improved and cite more recent measurements. Also address the threat-model
% realism concern: pre-registering many hallucinated package names is
% detectable; discuss feasibility/economics of slopsquatting at scale or cite
% observed in-the-wild instances. State upfront which LLMs are studied, and
% scope the generality claims (through Section 3.3, per Reviewer A) to
% verifiable domains rather than hallucination suppression in general.}

\section{Background}
\label{sec:background}

\subsection{Hallucinations in LLMs}
\label{subsec:hallucinations}

Hallucinations denote outputs that are linguistically fluent and superficially plausible but factually incorrect, unsupported by the input, or entirely fabricated. Their causes are well-studied~\cite{Ji_2023, lin-etal-2022-truthfulqa, wang-sennrich-2020-exposure}: training corpora contain noisy and contradictory claims that models absorb and reproduce confidently, the next-token prediction objective rewards fluency more directly than truthfulness, and exposure bias at decoding time compounds small errors as the model conditions on its prior mistakes. These factors make hallucinations persistent across tasks, domains, and model generations.

\subsection{Hallucination Reduction}
\label{subsec:reduction}

Existing mitigation approaches divide into \textbf{proactive} methods that modify the model and \textbf{reactive} methods that intervene at inference. Proactively, data curation and de-duplication during pretraining~\cite{lee2022deduplicating, longpre2024pretrainers} and RLHF~\cite{ouyang2022training} dominante. RLHF improves truthfulness on benchmarks like TruthfulQA but requires expensive annotation and struggles with rare failure modes; Constitutional AI~\cite{bai2022constitutional} cuts annotation cost via self-critique, but presupposes the model can detect its own errors, a capability that fails precisely when it hallucinates confidently.

Reactively, RAG~\cite{lewis2020retrieval} grounds responses in external knowledge bases at the cost of latency, architectural complexity, and the burden of maintaining the external corpus. DoLa~\cite{chuang2024dola} contrasts logits across transformer layers for 12-17\% gains on TruthfulQA without fine-tuning. Chain-of-thought~\cite{wei2022chainofthought} and self-consistency~\cite{wang2023selfconsistency} improve reasoning but not factuality directly, and incur multiple forward passes per query. Constrained decoding enforces structural correctness only and is limited to narrow output formats.
None of these enable \emph{post-deployment, targeted} removal of specific hallucination types: proactive methods require full retraining and alter global behavior, while reactive methods treat symptoms without touching the weights that produce them. \textit{Adaptive Unlearning} fills this gap.

\subsection{Unlearning in LLMs}
\label{subsec:unlearnign}

As LLMs are often trained on massive corpora, they may inadvertently memorize sensitive, private, or outdated information. \emph{Machine unlearning} removes or suppresses the influence of specific training data or knowledge in a trained model without retraining from scratch~\cite{nguyen2024surveyunlearning}, with applications to privacy compliance and post-hoc model correction. \emph{Exact unlearning} produces a model that behaves as if the target data had never been seen, but requires full retraining and is intractable at scale~\cite{nguyen2024surveyunlearning}. Recent work therefore focuses on empirical algorithms that fine-tune the model to forget targeted content efficiently~\cite{jang-etal-2023-knowledge}. 

One approach is \emph{gradient ascent}: fine-tuning the model by \emph{maximizing} loss on the data to forget, reducing its confidence in that completion. This has been successfully applied to LLMs~\cite{jang-etal-2023-knowledge}, but suffers from \emph{catastrophic collapse}. As the model is pushed away from certain knowledge, the overall performance of the model degrades. An overshooting optimizer can erase neighboring capabilities along with the target knowledge, often resulting in gibberish outputs~\cite{zhang2024negative}.

Negative Preference Optimization (NPO)~\cite{zhang2024negative}, inspired by DPO~\cite{rafailov2024directpreferenceoptimizationlanguage}, addresses this instability by structuring the objective as a preference over the target data rather than as raw error maximization, scaling unlearning to realistic dataset sizes without collapse. While the original NPO method used a frozen base model as a reference in the loss function, this causes the fine-tuned model to be compared to a stale base model after many epochs, a problem which reference-free variants have addressed~\cite{fan2025simplicityprevails}. Conservative alternatives such as logit suppression~\cite{huang2024offset} instead localize updates to preserve surface-level performance, but often achieve only partial unlearning. The model refrains from verbatim reproduction while still ``knowing'' the information, leaking it through paraphrase that can itself manifest as new hallucinations.

A separate line of work repurposes a known failure mode. Iteratively training a generative model on its own outputs induces \emph{model collapse}, a progressive loss of distributional diversity in which the tails of the output distribution are forgotten and the model degenerates into repetitive text~\cite{shumailov2024collapse, dohmatob2024strongmodelcollapse, bertrand2024stability}. The effect is irreversible. Further training on synthetic data cannot recover the lost diversity~\cite{gerstgrasser2024modelcollapse, shabgahi2025fortifaifendingrecursivetraining}. \emph{Partial Model Collapse} (PMC)~\cite{scholten2025pmc} triggers this process in a targeted way. It removes specific knowledge rather than degrading the model as a whole, and forms the methodological starting point for our approach in §\ref{sec:method}.

\subsection{Hallucination Reduction via Unlearning}
\label{subsec:viaUnlearning}

Recent work has begun to treat hallucinations as an unlearning target, removing the incorrect associations that produce it. The goal here is not privacy or regulatory compliance, but factual accuracy through forgetting incorrect knowledge or disabling spurious generation pathways. 

% However, a core challenge in using unlearning for hallucination reduction is \textbf{avoiding new hallucinations as a side-effect} in which a model inserts a different incorrect answer into the knowledge gap created through unlearning. 

Tan et al.~\cite{tan2025attention_shifting} characterize the central dilemma as a tradeoff between aggressive unlearning (which harms utility) and conservative unlearning (which leaves the model fluently filling the gap with new fabrications). They integrate refusal behavior into the unlearning objective, so the model responds with a safe fallback (e.g. "I'm not sure") rather than a new incorrect statement.

Despite extensive work on hallucination mitigation, only a small and recent subset of works casts hallucination suppression as an unlearning problem. Notably, most approaches in this direction have focused on multi-modal LLMs, leaving text-only hallucination via unlearning comparatively unexplored~\cite{xing-etal-2024-efuf, li2025analyzing}.

\section{Method}
\label{sec:method}

Adaptive Unlearning suppresses package hallucination through a closed-loop system. The loop iteratively discovers hallucination-inducing prompts, samples the model's outputs, and applies token-level unlearning to hallucinated completions. It extends PMC from a finite, enumerable forget set to the open-ended prompt space that hallucination reduction requires. Rather than removing discrete facts, AU repeatedly trains the model on its own hallucinated outputs to induce a controlled distributional collapse at hallucination-inducing contexts, while adaptive prompt mutation continuously supplies new instances so the model suppresses the pattern rather than memorizing prompt-output pairs.

% This is repititive a bit

% The key innovation is treating hallucinations not as discrete facts to remove but as emergent behaviors to suppress through targeted distribution collapse.

% AU operates on a fundamental insight: by repeatedly training a model on its own hallucinated outputs we can induce a controlled collapse of the model's distribution specifically for hallucination-inducing contexts. Unlike traditional PMC which assumes a fixed dataset of content to forget, AU continuously discovers new hallucination instances through adaptive prompt mutation, ensuring the model learns to suppress hallucination patterns rather than memorizing prompt-output pairs. 

\subsection{Problem Formulation}

Let $\mathcal{M}_\theta$ denote a language model with parameters $\theta$, and let $\mathcal{P}$ be a distribution over prompts that induce hallucinated packages. 

For $p \sim \mathcal{P}$, the model emits a token sequence $y = (y_1, \dots, y_T)$. A detection oracle $\mathcal{O}$ labels each package reference in $y$ as valid or hallucinated, yielding position sets $\mathcal{T}_{\text{forget}}$ and $\mathcal{T}_{\text{retain}}$, with all remaining positions forming $\mathcal{T}_{\text{reg}}$. We seek $\theta$ that suppresses generation at $\mathcal{T}_{\text{forget}}$ under prompt variation, preserves utility on general coding tasks, and generalizes to prompts unseen during training.

Standard unlearning assumes a forget set that is finite and known in advance. Here neither $\mathcal{P}$ nor the hallucinated spans it elicits can be enumerated ahead of time, and both shift as $\theta$ changes. Both must be discovered during training, which is the requirement that motivates the closed loop in \S\ref{subsec:nested} and \S\ref{subsec:mutation}.

% This differs fundamentally from standard unlearning where the forget set is finite and known apriori. In hallucination reduction, both $\mathcal{P}$ and the specific hallucinations $h$ are \textit{potentially infinite} and must be discovered during training.

\subsection{Loss Functions}
\label{subsec:loss_functions}

AU's objective combines three terms. Cross-entropy (CE) reinforces tokens the oracle validates. NPO suppresses tokens it rejects, and a scaled cross-entropy term anchors the remaining tokens to the base distribution. The tri-mask in \S\ref{subsec:au_loss} ensures all three terms act on disjoint token subsets. AU extends PMC~\cite{scholten2025pmc} from finite to open-ended forget sets, and we summarize PMC's formal objective in Appendix~\ref{app:pmc_formula}.

The cross-entropy loss function applies standard next-token prediction:
\begin{equation}
\mathcal{L}_{\text{CE}}(\theta) = -\sum_{\{t \,:\, m_t = 1\}} \log p_\theta(y_t \mid y_{<t}, x)
\end{equation}

where $x$ is the input prompt and $y_{<t}$ is the preceding token sequence. This objective reinforces valid package tokens and structural code elements, serving as the primary utility-preserving component of the hybrid loss.

NPO \cite{zhang2024negative} frames unlearning as preference optimization, providing more stable gradients than gradient ascent. We apply it at token granularity,
\begin{equation}
\mathcal{L}_{\text{NPO}}(\theta) = -\mathbb{E}_{y^-}\left[
\log \sigma \left(
-\beta \log \frac{\pi_\theta(y^- \mid p)}{\pi_{\text{ref}}(y^- \mid p)}
\right)
\right]
\end{equation}

where $y^-$ are hallucinated completions, $\pi_{\text{ref}}$ is the frozen base model, $\beta$ is a temperature controlling suppression strength, and $\sigma$ is the sigmoid function.

\begin{figure*}[!t]
  \centering
  \includegraphics[width=0.9\textwidth]{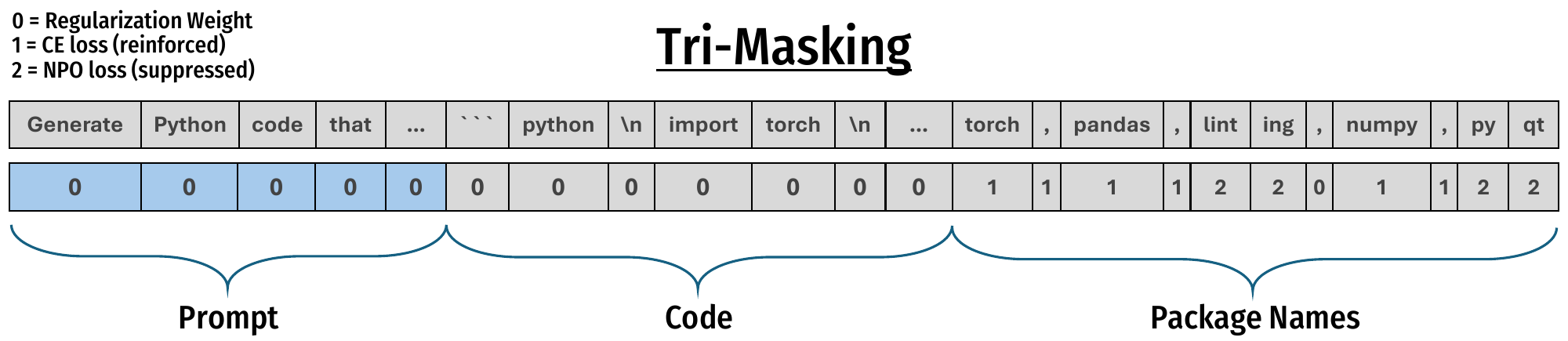}
  \caption{Token-level tri-masking applies targeted loss functions across every sample for precise suppression and reinforcement.}
  \label{fig:tri_masking}
\end{figure*}

\subsection{Tri-Masking}
\label{sec:tri-masking}

Tri-masking routes gradients at the token level, determining which positions are reinforced, suppressed, or held near the base distribution. Prior masking approaches in unlearning are binary. Tri-masking adds a third category, and the resulting mask $\mathbf{m} = (m_1, \dots, m_T) \in \{0,1,2\}^T$ assigns one value to every token position.

% \textbf{Regularize ($m_t = 0$).} All positions that are not package references, including the prompt, surrounding code, and punctuation. These receive cross-entropy scaled by $\lambda_{\text{reg}}$, anchoring the model at non-target positions and preventing the parameter drift that otherwise degrades output quality. \textbf{Reinforce ($m_t = 1$).} Package tokens the oracle confirms exist, which receive standard cross-entropy. \textbf{Suppress ($m_t = 2$).} Package tokens the oracle rejects, which
% receive NPO suppression gradients.

\begin{itemize}
    \item \textbf{Mask = 0 (Regularize):} Tokens receive a regularization weight in the form of cross-entropy loss scaled by $\lambda_{\text{reg}}$. This is applied to all tokens that are not package names, including prompts, code, and punctuation. This has the effect of anchoring a model's distribution at non-target positions, preventing parameter drift that otherwise causes output degeneration. 
    \item \textbf{Mask = 1 (Reinforce):} Tokens receive standard reinforcement through cross-entropy loss. Applied to correct tokens, which in our use case are the valid package names. %Reinforcing structural elements and EOS tokens are critical in preserving natural output and preventing the model getting stuck in a repetitive loop.
    \item \textbf{Mask = 2 (Suppress):} This applies to hallucinated tokens which receive suppression gradients via NPO loss. In our use case this represents hallucinated package names. 
\end{itemize}

The tri-mask $\mathbf{m} = (m_0, m_1, m_2)$ is generated dynamically for each training sample based on the domain-specific hallucination detection. For package hallucinations, our detection pipeline identifies non-existent package names and marks those tokens with mask = 2, marks valid packages with mask = 1, and marks all other tokens with mask = 0. It is important to note that this tri-masking process is only necessary during the fine-tuning of the model and is not required during regular model use. 

A critical finding on contextual and punctuation tokens was discovered as our initial implementation assigned no gradient to any non-package token. This led to a severe degradation in the model's coding ability. We then implemented a regularization term that anchored these critical tokens to the base model, preventing incoherent responses and repetitive over-generation, where the model continued producing tokens indefinitely.

\subsection{The Adaptive Unlearning Objective}
\label{subsec:au_loss}

The full Adaptive Unlearning objective combines the cross-entropy and NPO losses introduced above through the tri-mask, applying each to a disjoint token population within the same sequence. Concretely, for a tokenized sequence $y = (y_1, \dots, y_T)$ with tri-mask $\mathbf{m} = (m_1, \dots, m_T) \in \{0, 1, 2\}^T$, define the partitioned token-position sets
\begin{align}
\mathcal{T}_{\text{reg}}    &= \{ t : m_t = 0 \}, \\
\mathcal{T}_{\text{retain}} &= \{ t : m_t = 1 \}, \\
\mathcal{T}_{\text{forget}} &= \{ t : m_t = 2 \}.
\end{align}
The Adaptive Unlearning objective is then:
\begin{equation}
\label{eq:au_loss}
\mathcal{L}_{\text{AU}}(\theta) \;=\;
\lambda_{\text{retain}} \, \mathcal{L}_{\text{CE}}^{\,\mathcal{T}_{\text{retain}}}(\theta)
\;+\;
\lambda_{\text{forget}} \, \mathcal{L}_{\text{NPO}}^{\,\mathcal{T}_{\text{forget}}}(\theta)
\;+\;
\lambda_{\text{reg}}    \, \mathcal{L}_{\text{CE}}^{\,\mathcal{T}_{\text{reg}}}(\theta),
\end{equation}

where each term is the corresponding loss from \S\ref{subsec:loss_functions} restricted to the indicated token positions.

The retain term $\mathcal{L}_{\text{CE}}^{\,\mathcal{T}_{\text{retain}}}$ inherits PMC's self-training mechanic. Rather than reinforcing externally curated ground-truth tokens, it reinforces the valid package tokens the current model $\pi_\theta$ already emits and that $\mathcal{O}$ confirms. Iteratively training on these self-generated, oracle-validated tokens drives the conditional distribution on package-recommendation prompts toward the outputs the model emits \emph{and} that survive verification. 

This is a per-token specialization of PMC's curated-self-training~\eqref{eq:pmc_objective}, with curation operator $\mathcal{C}$ instantiated by the oracle $\mathcal{O}$ filter rather than a global preference function. The forget term $\mathcal{L}_{\text{NPO}}^{\,\mathcal{T}_{\text{forget}}}$ pushes probability mass off the complementary subset, meaning tokens the model produces but the oracle rejects. The regularization term $\mathcal{L}_{\text{CE}}^{\,\mathcal{T}_{\text{reg}}}$ anchors the model on the remaining structural and contextual tokens near the base model.

A partition contributes only when non-empty. A completion containing only hallucinated packages leaves $\mathcal{T}_{\text{retain}} = \emptyset$ and drops that term for the sample, and the same holds for the other two. One objective therefore covers forget-only, retain-only, and mixed sequences without special-casing.

% Each partition contributes only when non-empty: if $\mathcal{T}_{\text{retain}} = \emptyset$ for a given sample (e.g., a query that produces only hallucinated package names), the corresponding term is dropped from~\eqref{eq:au_loss} for that sample, and analogously for the other two partitions. This is the mechanism by which the same objective handles the full range of sample compositions encountered during training: forget-only sequences (NPO-only updates), valid-only sequences (CE-only updates), and mixed sequences (both updates simultaneously on disjoint tokens).

\paragraph{Interaction with the tri-mask.} Disjointness is what makes~\eqref{eq:au_loss} well-posed. Because the three terms act on non-overlapping token positions, the gradients from $\mathcal{L}_{\text{CE}}^{\,\mathcal{T}_{\text{retain}}}$ (pushing probability mass toward valid package tokens) and $\mathcal{L}_{\text{NPO}}^{\,\mathcal{T}_{\text{forget}}}$ (pushing probability mass away from hallucinated package tokens) cannot conflict at a single token. Applying CE and NPO sequence-wide instead produces opposing gradients on the same logit vector at the same step, the mechanism behind the catastrophic-collapse instability documented for naïve gradient ascent~\cite{zhang2024negative}.

% Reviewrs wanted something like this
Disjointness is an assumption. It holds for package identifiers, which sit in delimited import positions and rarely influence the surrounding code. It weakens when a hallucinated name collides with a legitimate identifier, or shares subword pieces with a valid name so that suppression leaks. It fails entirely for targets that spread incorrectness across a
clause rather than a delimited span, which we treat as out of scope.
\subsection{Nested Training Loops}
\label{subsec:nested}

PMC-based unlearning cannot run on a fixed dataset. Inducing collapse requires regenerating model outputs as $\theta$ changes. However, regenerating samples every epoch is unstable, because the model does not register a given sample often enough to consolidate which patterns to suppress. AU therefore separates resampling from optimization across two timescales.

\textbf{Outer Epochs ($N_{outer}$)} control resampling frequency. At the start of each, the current model generates fresh completions for all active prompts in inference mode, and the detector labels them to build tri-masks. This tracks the distribution as it shifts under training.

\textbf{Inner Epochs ($N_{inner}$)} enable knowledge consolidation. The model trains repeatedly on the cached completions from the current outer epoch, letting gradients accumulate and reinforce the suppression pattern before the samples are regenerated. This prevents the usual instability where the model constantly "chases" shifting samples.

This design choice is critical for hallucination reduction. In ablation studies, training without inner epochs (equivalent to $N_{inner} = 1$) failed to reduce hallucination rates. Hallucination rate fell monotonically at $N_{\text{inner}} \geq 10$.

\subsection{Prompt Mutation Strategy}
\label{subsec:mutation}

% A fundamental challenge in hallucination reduction is generalization: the model must learn to suppress hallucination patterns, not merely memorize that specific prompts should not produce hallucinations. Static prompts risk memorization, where the model learns ``don't hallucinate for these exact prompts" but fails to generalize for variations.

Suppression must generalize beyond the prompts it was trained on. A fixed prompt set risks teaching the model to avoid hallucinating on those exact inputs while leaving the underlying behavior intact.

AU mutates prompts as they stop producing signal. A prompt is \emph{exhausted} when $N$ sampled completions contain no hallucinated package, or when it has been trained for five outer epochs, whichever comes first. The model is then shown the exhausted prompt and instructed to write a reworded version of it $p' = \textsc{Mutate}(p)$, always from the original seed rather than the latest variant so that mutations do not compound. The rewrite is constrained to preserve the task's library requirements and technical goal while varying the phrasing. The variant should elicit different code, but code requiring the same packages. Mutation resets the exhaustion state and clears the cached code. This forces the model to generalize prior training iterations to name valid packages and suppress hallucinated ones rather than just memorizing the original wording.

% This chained mutation approach ensures continuous discovery of new hallucination-inducing contexts. Our results demonstrate strong generalization: after training on original prompts and 6 rounds of mutations, the model achieved \textbf{3.9}\% hallucination rate on a set of completely unseen prompts, compared to \textbf{12.7}\% baseline without mutation.

After six rounds of mutation, hallucination rate on held-out prompts was 3.9\%, against 12.7\% for training on the original prompts alone.

% Adaptive mutation implements a form of curriculum learning where the model progressively encounters harder examples. By mutating prompts when they stop inducing hallucinations, we ensure the training distribution stays at the ``edge" of the model's capabilities, maximizing learning signal while preventing both trivial examples (already suppressed) and impossible examples (completely novel patterns).

\subsection{Training}

The complete AU training procedure integrates all components, as seen in Fig.~\ref{fig:pipeline}. This procedure ensures that: (1) the model encounters diverse hallucination-inducing contexts through mutation, (2) training is stable through nested epochs, (3) gradient flow is precise through tri-masking, and (4) both suppression and preservation occur through the mixed loss objective.

\begin{figure}[!t]
  \centering
  \includegraphics[width=0.45\textwidth]{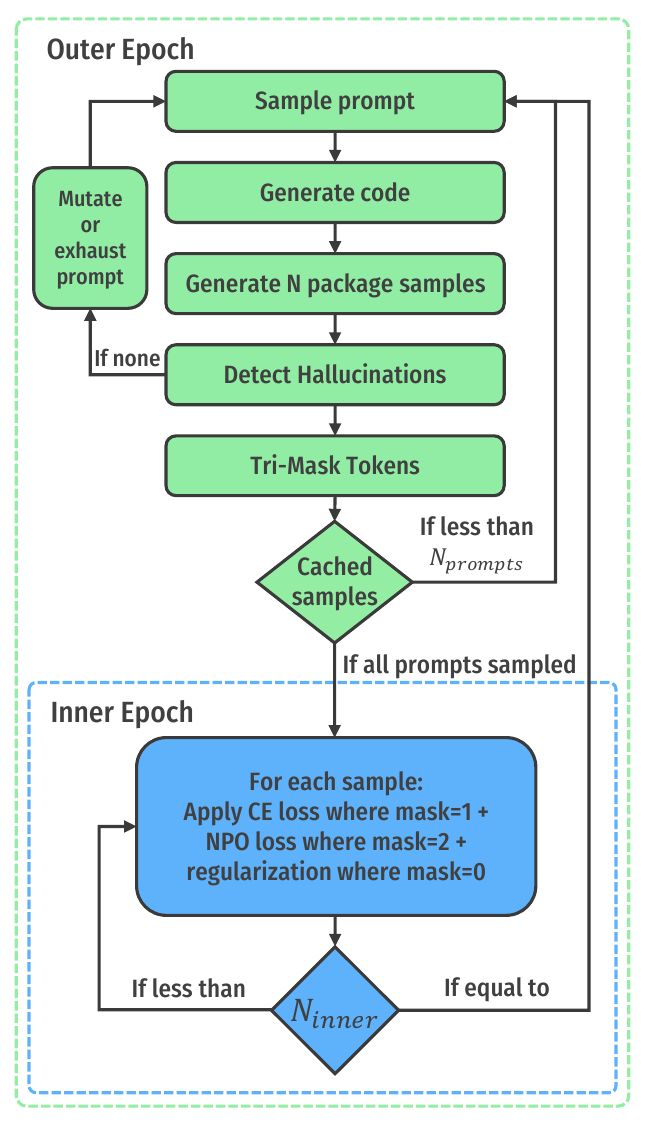}
  \caption{\textbf{Nested outer/inner epoch structure.} Nested outer/inner epoch structure. Outer epochs regenerate samples and invoke prompt mutation; inner epochs train on cached samples via the tri-masked hybrid objective, stabilizing the unlearning loop. This process is detailed in \S\ref{subsec:nested}.}
\label{fig:nested_epochs}
  \label{fig:training}
\end{figure}

\section{Experimental Evaluation}
\label{sec:experiment}

% We evaluate AU on package hallucination in code generation, a setting that combines verifiable ground truth with direct security consequences. We first motivate the choice of domain (\S\ref{sec:domain}), describe our detection pipeline (\S\ref{sec:detection}) and experimental configuration (\S\ref{sec:setup}), and then define the baselines and evaluation protocol used throughout (\S\ref{sec:baselines}--\S\ref{sec:protocol}).
We evaluate Adaptive Unlearning on package hallucination in code
generation, which pairs verifiable ground truth with a live attack surface.

\subsection{Domain and Detection}
\label{sec:domain}
Package hallucinations arise when a code-generating model emits references to libraries absent in the target ecosystem's registry~\cite{spracklen2025}. A model asked for a Python plotting script may import the legitimate \texttt{matplotlib} alongside a fabricated \texttt{quickplot}. This failure mode is particularly dangerous for four reasons:

\begin{itemize}
    \item \textbf{Syntactic validity.} Hallucinated imports are syntactically well-formed and indistinguishable from legitimate ones at the parser level.
    \item \textbf{Semantic plausibility.} Fabricated names frequently follow natural naming conventions (e.g., \texttt{fastjson}, \texttt{easyplot}) and appear credible to human reviewers.
    \item \textbf{Attack surface.} Hallucinated package names enable \emph{package confusion} and \emph{slopsquatting} attacks~\cite{spracklen2025}. an adversary who observes (or predicts) hallucinated names can publish malicious packages under those names, which are then available to be installed and executed by downstream users of the model.
    \item \textbf{Misplaced trust.} As developers increasingly adopt model-generated code with limited review, hallucinated dependencies provide a low risk/high reward injection vector into otherwise legitimate software supply chains.
\end{itemize}

% Package hallucination is a particularly suitable evaluation domain for unlearning methods because it satisfies three properties that are rarely available jointly:

% \textbf{(i) Verifiable ground truth.} Unlike open-domain factual hallucinations, package existence is a deterministic property: a name is either resolvable against the PyPI index or it is not.

% \textbf{(ii) Open-ended prompt space.} Arbitrarily many natural-language specifications can elicit package references, varying in task, domain, style, and complexity.
% \textbf{(iii) Operational relevance.} Package hallucination is an actively exploited vulnerability in deployed code assistants, making the evaluation both methodologically sound and directly tied to a real-world threat model.

The domain also has properties that are rare for unlearning evaluation. Package existence is deterministic, since a name either resolves against the PyPI index or does not. The prompt space is open-ended, as arbitrarily many specifications elicit package references. And the failure mode is operationally live rather than hypothetical.

We detect hallucinations by cross-referencing outputs against a canonical PyPI snapshot. Given a prompt $p$, the model produces completion $y$. We parse $y$ for explicit \texttt{pip install} directives, then separately query the model for the packages required to run $y$ and parse the response $r$ for identifiers. Each identifier is resolved against the snapshot, and
we emit the token-level mask $\mathbf{m}$ over $r$ as described in \S\ref{sec:tri-masking}. A visual description of the pipeline is in Appendix \ref{app:app_detection}.

% \subsection{Hallucination Detection Pipeline}
% \label{sec:detection}

% We detect package hallucinations via an automated pipeline that cross-references model outputs against a canonical snapshot of the PyPI index:

% \begin{enumerate}
%     \item \textbf{Code generation.} Given a coding prompt $p$, the model produces a completion $y$.
%     \item \textbf{Explicit-install extraction.} We parse $y$ to extract any explicit \texttt{pip install} directives and their arguments.
%     \item \textbf{Dependency elicitation.} We additionally query the model for the set of packages required to execute $y$, yielding a response $r$. We parse $r$ for package identifiers and cross-reference each against the PyPI snapshot.
%     \item \textbf{Tri-mask construction.} We emit a token-level mask $\mathbf{m}$ over $r$ as described in \S~\ref{sec:tri-masking}.
% \end{enumerate}

\subsection{Setup}
\label{sec:setup}

\paragraph{Models.} 
% We evaluate on two instruction-tuned, code-specialized models from the DeepSeek-Coder family~\cite{guo2024deepseekcoderlargelanguagemodel}:
% \begin{itemize}
%     \item \texttt{DeepSeek-Coder-7b-instruct-v1.5}\footnote{\url{https://huggingface.co/deepseek-ai/deepseek-coder-7b-instruct-v1.5}} (DeepSeek-7B), a \\
%     dense 7B-parameter model.
%     \item \texttt{DeepSeek-Coder-V2-Lite-Instruct}\footnote{\url{https://huggingface.co/deepseek-ai/DeepSeek-Coder-V2-Lite-Instruct}} (DeepSeek-16B), a Mixture-of-Experts model with 16B total parameters and approximately 2.4B active parameters per token.
% \end{itemize}

We evaluate two instruction-tuned models from the DeepSeek-Coder family~\cite{guo2024deepseekcoderlargelanguagemodel}, \texttt{DeepSeek-Coder-7b-instruct-v1.5} (dense, 7B) and \texttt{DeepSeek-Coder-V2-Lite-Instruct} (MoE, 16B total and roughly 2.4B
active per token). The pair spans dense and sparsely activated architectures at a scale comparable to prior unlearning work~\cite{scholten2025pmc, zhang2024negative}. All runs use
full-parameter fine-tuning rather than low-rank adaptation, so results reflect the objective rather than adapter capacity. Experiments ran on AMD MI210 and MI300 GPUs.

% The pair spans dense and sparsely activated architectures and represents a meaningful step up in scale relative to prior unlearning work in this space: the original PMC evaluation~\cite{scholten2025pmc} reports results on 3B and 12B models, and NPO~\cite{zhang2024negative} is evaluated at the 7B scale, whereas our setup covers both a dense 7B model directly comparable to NPO's setting and a 16B MoE configuration that tests whether the same unlearning machinery composes with the sparse-routing dynamics that increasingly characterize frontier-scale code models. All Adaptive Unlearning runs use \textbf{full-parameter fine-tuning}: we update every trainable parameter rather than adopting parameter-efficient methods such as LoRA, ensuring that the reported results reflect the full capacity of the unlearning objective rather than the expressive limits of a low-rank adapter. Experiments are conducted on AMD MI210 and MI300 GPUs.

\paragraph{Dataset.} We curate a set of 20 code-generation prompts drawn from the benchmark introduced by Spracklen et al.~\cite{spracklen2025}. Prompts are selected subject to two criteria: (i) under the base model's initial sampling, the completion contains at least one valid and one hallucinated package, ensuring the prompt is a meaningful stressor; and (ii) the set collectively covers a diverse range of programming tasks and application domains, mitigating topical bias. We intentionally selected prompts that elicit a large number of hallucinations to thoroughly test our method, which explains why our baseline results being much larger than in the original study. For testing, we used ChatGPT to develop 2 variants of each initial coding prompt. This 40 prompt test set is completely held-out of training and enables generalized testing on unseen prompts. The full list of initial coding prompts can be found in Appendices~\ref{app:app_prompts} and~\ref{app:app_initial_prompts}.

\paragraph{Configuration.} Batch size 8, maximum sequence length 1024, six mutations per prompt, five cached samples per outer epoch, and exhaustion at five outer epochs or zero hallucinations. Loss weights are $(\lambda_{\text{retain}}, \lambda_{\text{forget}}, \lambda_{\text{reg}}) = (1.00, 1.25, 1.00)$, selected over a small grid reported in
Appendix~\ref{app:hyperparams}. All methods share the same generation, parsing, and detection pipeline, and differ only in the loss applied and whether samples are regenerated online.

\subsection{Baselines and Ablations}
\label{sec:baselines}
We organize our experimental comparisons along two axes. The \textbf{baselines} situate Adaptive Unlearning against the prior art in hallucination suppression and unlearning. The \textbf{ablated variants} isolate the contribution of individual components of our method by zeroing them out.

\paragraph{Methods.}
\begin{enumerate}
    \item \textbf{Base.} The unmodified DeepSeek-Coder model, establishing the reference hallucination rate and unmodified coding performance.
    \item \textbf{Gradient Ascent (GA).} Direct gradient ascent on the negative log-likelihood of hallucinated tokens, the canonical first-line unlearning baseline.
    \item \textbf{Negative Preference Optimization (NPO).} The preference-based unlearning objective of Zhang et al.~\cite{zhang2024negative}, applied to suppress hallucinated tokens.
    \item \textbf{PMC.} The self-training distribution-collapse method of Scholten et al.~\cite{scholten2025pmc}, which fine-tunes on the model's own generations selected against a preference function.
    \item \textbf{Adaptive Unlearning (ours).} The full method: a hybrid token-level objective routed through a tri-mask, embedded in an adaptive prompt-mutation loop with nested-epoch sample caching.
\end{enumerate}

\paragraph{Ablated variants.}
To attribute AU's performance to its individual components, we evaluate two ablations that zero out one loss term while holding the rest of the method fixed:

\begin{itemize}
    \item \textbf{AU CE-Only} ($\lambda_{\text{retain}}=1.0$, $\lambda_{\text{forget}}=0$): the NPO suppression term is removed; the model receives reinforcement gradients on valid package tokens but no suppression on hallucinated ones.
    \item \textbf{AU NPO-Only} ($\lambda_{\text{forget}}=1.0$, $\lambda_{\text{retain}}=0$): the CE reinforcement term is removed; the model receives suppression gradients on hallucinated tokens but no reinforcement of valid ones.
\end{itemize}

We additionally examine the sensitivity of AU to two key hyperparameters of the adaptive scaffold: the number of inner epochs over cached samples, and the maximum number of prompt mutations per prompt. We return to all comparisons in \S\ref{sec:ablations}.

% We additionally conduct ablation experiments on two key hyperparameters:
% \begin{enumerate}
%     \item Inner epochs
%     \item Number of mutations
% \end{enumerate}

% We return to these comparisons in \S\ref{sec:ablations}.

\subsection{Metrics}
\label{sec:protocol}
Following Spracklen et al.~\cite{spracklen2025}, each prompt receives two follow-up queries, one asking which packages are required to run the code and one asking which would be useful for the task. The first targets strict runtime dependencies and the second elicits optional suggestions. 
\begin{description}
    \item [\textbf{Mode 1 (required).}] ``Which packages are required to run this code?'' Targeting strict runtime dependencies.
    \item [\textbf{Mode 2 (helpful).}] ``What packages would be useful to solve this task?'' Eliciting optional and exploratory suggestions.
\end{description}
Our primary metric is hallucination rate, $\text{HR} = N_{\text{halluc}} /
N_{\text{total}}$, over all identifiers emitted across both modes.

We measure utility on EvalPlus~\cite{evalplus}, whose expanded HumanEval+ and MBPP+ suites~\cite{chen2021codex, austin2021programsynthesislargelanguage} catch solutions that pass the original tests but fail on edge cases. We report pass@1 against the unmodified base model. To bound collateral damage, we also report KL-divergence from the base model on three response
types, namely code samples, instruction following, and package recommendations. An ideal method isolates its updates to the package context.
\newcommand{\method}[1]{\textsc{#1}}
\newcommand{\best}[1]{\textbf{#1}}
\newcommand{\secondb}[1]{\underline{#1}}

\section{Results}
\begin{table*}[h]
\centering
\small
\setlength{\tabcolsep}{4pt}
\begin{tabular}{l cc ccc ccccc}
\toprule
 & \multicolumn{2}{c}{\textbf{Hallucination Rate (\%)}}
 & \multicolumn{3}{c}{$D_{\mathrm{KL}}(\pi_{\theta}\,\|\,\pi_{\text{baseline}})$ }
 & \multicolumn{5}{c}{\textbf{Coding Benchmarks (pass@1)} $\uparrow$} \\
\cmidrule(lr){2-3} \cmidrule(lr){4-6} \cmidrule(lr){7-11}
\textbf{Variant}
 & \textbf{Total $\downarrow$} & \textbf{Abs. Reduction $\uparrow$}
 & \textbf{Code$\downarrow$} & \textbf{Instr.$\downarrow$} & \textbf{Pkg. Avg.$\uparrow$}
& \textbf{HE} & \textbf{HE+} & \textbf{MBPP} & \textbf{MBPP+} & \textbf{Avg.}\\
% \midrule
% \multicolumn{11}{l}{\textit{DeepSeek-1B}} \\
% \midrule
% Base                & 34.53 & 31.16 & 32.97 & -- & -- & -- & -- & 63.4 & 58.5 & 60.3 & 51.9 \\
% GA                  & 9.69 & 14.74 & 12.03 & -- & -- & -- & -- & 58.5 & 53.7 & 56.6 & 48.1 \\
% NPO                 & 11.64 & 6.6 & 9.24 & -- & -- & -- & -- & 62.8 & 57.3 & 55.3 & 47.6 \\
% PMC                 & 23.64 & 33.33 & 28 & -- & -- & -- & -- & -- & -- & -- & -- \\
% Adaptive Unlearning & \best{5.58} & \best{1.23} & \best{3.71} & -- & -- & -- & -- & 53.7 & 49.4 & 52.6 & 46.6 \\

\midrule
\multicolumn{11}{l}{\textit{DeepSeek-7B}} \\
\midrule
Base                 & 21.23 & {0} & -- & -- & -- & \secondb{66.5} & \secondb{62.22} & {74.1 }& \best{65.1} & \secondb{66.98} \\
GA                  & \secondb{7.2 }& \secondb{14.03} & \best{0.0031} & \best{0.0679} & {0.1529} & \secondb{66.5} & 61.6 & \best{74.6} & \best{65.1} & 66.95\\
NPO                 & 14.17 & 7.06 & \best{0.0191} & {0.3565} & \secondb{0.7794} & {65.9} & 59.8 & {74.2} & \secondb{64.3} & {66.07}\\
PMC                 & 11.84 & 9.39 & \secondb{0.0060} & \secondb{0.1820} & 0.3622 & 59.8 & 56.1 & 48.9 & 41.8 & 51.65\\ 
AU (Ours) & \best{2.56} & \best{18.67 }& 0.0796 & 0.9905 & \best{1.4211} & \best{73.2} & \best{64.6} & \secondb{74.3} & {62.2} & \best{68.57} \\

\midrule
\multicolumn{11}{l}{\textit{DeepSeek-16B}} \\
\midrule
Base                & 27.75 & 0 & -- & -- & -- & \best{81.1} & \best{75} & 83.1 & \secondb{70.4} & \best{77.4} \\
GA                  & 13.75 & \secondb{14} & \secondb{0.0301} & 0.3236 & \best{1.6029} & \secondb{79.9} & \best{75} & \best{83.9} & \best{70.6}  & \secondb{77.35} \\
NPO                 & 17.58 & 10.17 & \best{0.0298} & \best{0.2691} & \secondb{1.247} & 78 & \secondb{73.8} & \secondb{83.3} & 69.9 & 76.17 \\
PMC                 & 16.66 & 11.09 & 0.0521 & 0.2997 & 0.8683 & 72.6 & 66.5 & 71.4 & 60.8 & 67.82 \\
% Adaptive Unlearning & 10.07 & 0.0479 & 0.457 & 0.9016 & 80.5 & 75.6 & 82.3 & 69.8 & 77.05 \\
% Adaptive Unlearning & 9.25 & 0.0465 & 0.3033 & 0.5595 & 79.9 & 75 & 82.8 & 69.3 & 76.75 \\
% Adaptive Unlearning & 6.75 & 0.0361 & 0.3117 & 0.7037 & 76.2 & 70.7 & 79.6 & 68 & 73.62 \\
AU (Ours) & \best{6.13} & \best{21.62} & 0.0368 & \secondb{0.299} & 0.6782 & 73.2 & 69.5 & 81.2 & 68.8 & 73.17 \\

\bottomrule
\end{tabular}
\vspace{2pt}
\caption{The main results table includes the total hallucination rate as a percentage (lower is better) as well as the absolute difference in hallucination rate compared to baseline (higher is better). The KL-divergence results are shown for each of the coding, instruction-following, and package recommendation queries, measured in nats. The last five columns document code benchmark scores, measured at pass@1 rate. Findings from this table are discussed in detail in \S\ref{sec:main_results}. Best is in bold and second-best is underlined, per column.}
\label{tab:main_results}
\end{table*}

\begin{figure*}[!t]
  \centering
  \includegraphics[width=0.99\textwidth]{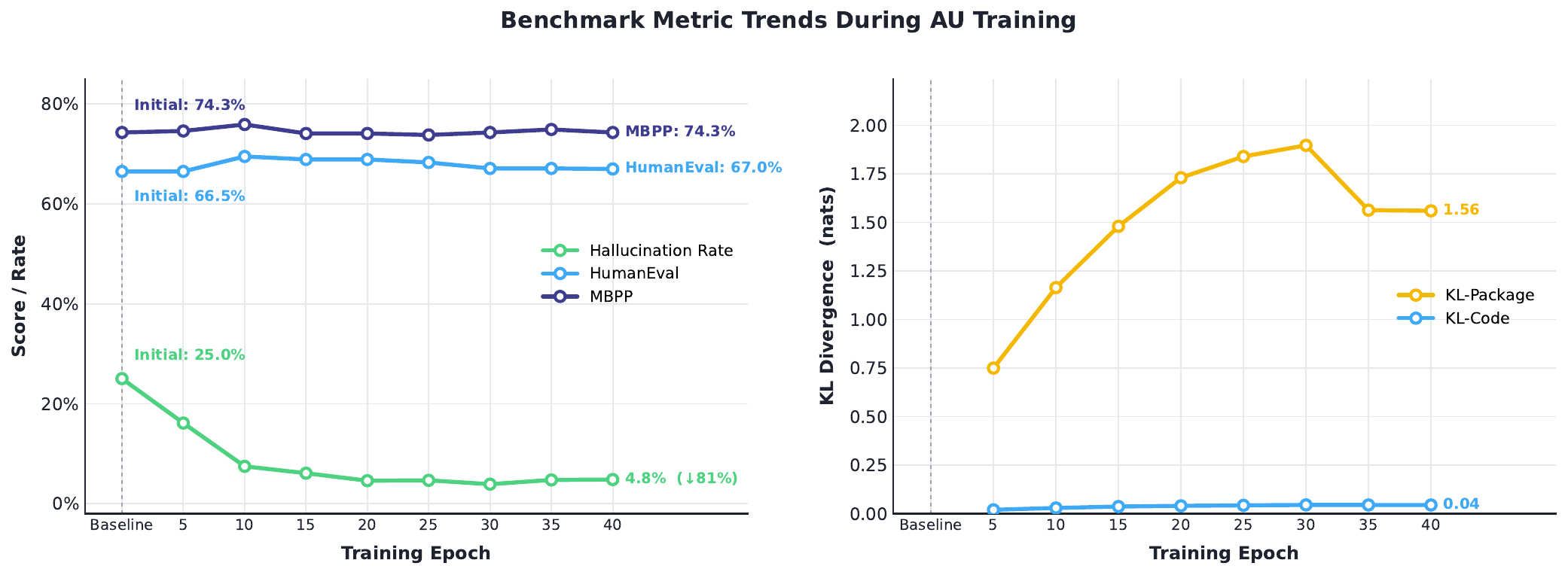}
  \caption{AU training metrics across 40 epochs. Hallucination rate drops \textbf{88\%} while coding benchmarks remain stable. KL-divergence is concentrated in the package distribution, confirming surgical targeting.} %AU training metrics across 40 epochs of unlearning. Left: Package hallucination rate falls from 25\% to 4.8\%, an 81\% reduction, while coding benchmarks remain stable, confirming that targeted suppression does not erode general utility. Right: KL-divergence from the baseline model is concentrated in the package name distribution (yellow) while the broader code distribution is essentially unchanged (blue). Together, these trends demonstrate that AU's intervention is surgical: the model's behavior shifts precisely where hallucinations live but not beyond.}
  \label{fig:trends}
\end{figure*}

A useful post-deployment package hallucination mitigation method must do three things at once. It should reduce the hallucination rate, leave coding ability intact, and confine distributional change to the package generation rather than perturbing model broadly. A method that suppresses hallucinations at the cost of code quality has relocated the failure, not fixed it. Table \ref{tab:main_results} and Figure \ref{fig:trends} therefore report all three axes together for every method.

We show that AU beats prior baselines on this tradeoff, quantify its cost in utility and drift, and attribute the effect to individual components (\S\ref{sec:ablations}).

\subsection{Primary Findings}

Table~\ref{tab:main_results} reports the cross-method comparison. \textbf{Adaptive Unlearning achieves the strongest overall tradeoff among the evaluated baselines with the lowest package hallucination rates on both models}. AU is the only method that simultaneously drives hallucination rates down to low single digits, preserves strong coding performance, and confines its largest behavioral changes to the package-generation setting rather than ordinary code generation. This is the correct lens for evaluating a post-deployment mitigation method. Reducing hallucinations alone is not enough if the model’s coding ability collapses, and preserving utility alone is not enough if the model continues to emit exploitable fabricated package names.

% On that combined criterion, AU is the clear winner.

AU reaches the lowest hallucination rate on both models, 2.56\% on DeepSeek-7B and 6.13\% on DeepSeek-16B, from base rates of 21.23\% and 27.75\%. On 7B this is an 88\% relative reduction, and a 64\% reduction against GA, the strongest non-AU baseline at 7.2\%. All testing is executed on a held-out test set of prompts that were unseen during training.

% On the primary security metric, AU reduces total hallucination rate from 21.2\% in the base DeepSeek-7B model to 2.5\%, compared to 7.2\% for GA, 14.17\% for NPO, and 11.8\% for PMC. Relative to the base model, this in a 87.9\% reduction in total hallucinations and a 64\% reduction relative to the strongest non-AU baseline, GA. 

Package hallucination suppression does not come from broad capability loss. On 7B model, AU attains the highest coding average in the table at 68.5\%, 1.5\% higher than the GA and NPO baselines, improving on three of four benchmarks with a small drop on MBPP+.PMC shows the alternative outcome. It cuts hallucination rate to 11.84\% while its coding average falls from 66.98\% to 51.65\%, which is the failure mode the joint view exists to expose. The model trained via AU has clearly retained its coding proficiency while delivering state-of-the-art hallucination reduction performance.

AU's results on DeepSeek-16B, \textbf{while still the strongest among all evaluated methods}, show a wider gap in utility suppression than on DeepSeek-7B. Package hallucination rate drops to 6.1\% (compared to 2.5\% on 7B) at the cost of a 4.2\% decline in average coding benchmark score (where 7B actually improved). We attribute the difference to MoE routing. With roughly 2.4B of 16B parameters active per token, each expert receives a sparse and fragmented unlearning signal, which makes consolidation harder and raises the risk of collateral updates to experts handling unrelated capabilities. A dense model updates every parameter on every forward pass. We do not test this mechanism directly and offer it as a hypothesis.

Distribution drift concentrates where AU targets it. On 7B model, AU induces 0.0796 KL on code against 1.4211 on package queries, \textbf{a ratio near 18 to 1}, with 0.9905 on instruction following.. AU is not the most conservative update in absolute terms where GA and PMC both perturb code less. However, the relevant property in this context is the \textit{ratio}. The model is aggressive in pursuing inconsistencies in package inducing settings, providing strong updates to that specific part of the distribution while leaving the coding distribution relatively untouched. This is exactly the form of targeted change a practical post-deployment mitigation should aim for

% Drift concentrates where AU targets it. On 7B, AU induces 0.0796 KL on code against 1.4211 on package queries, a ratio near 18 to 1, with 0.9905 on instruction following. AU is not the most conservative update in absolute terms, and GA and PMC both perturb code less. The relevant property is the ratio. AU spends its update budget rewriting package-recommendation
% behavior rather than degrading ordinary code generation. Figure~\ref{fig:trends} shows the same asymmetry over training. Hallucination rate falls sharply and flattens while HumanEval and MBPP hold steady, and KL rises in the package distribution while staying near
% baseline for code.

Figure \ref{fig:trends} complements Table \ref{tab:main_results} by showing that AU reaches this endpoint through a \textbf{stable trajectory} rather than an erratic or destructive one. In the left panel, package hallucination rate drops sharply early in training and then flattens, while HumanEval and MBPP remain essentially stable throughout. The right panel shows the same asymmetry distributionally. KL divergence stays near baseline for ordinary code generation and rises mainly in the package-related distribution. 

Taken together, Table \ref{tab:main_results} and Figure \ref{fig:trends} establish the main claims of the paper. AU is neither the most aggressive hallucination suppressor nor the most conservative fine-tuning procedure. It is the one that balances all three requirements, cutting package hallucination rate substantially while preserving coding utility and confining the update to package generation.

% Added the explanation into the 16B paragraph as we basically ignored it in the text
% \paragraph{Architecture-dependent utility tradeoff.} AU's results on DeepSeek-16B, while still the strongest among all evaluated methods, show a wider gap in utility suppression than on DeepSeek-7B: hallucination rate drops to 6.1\% (vs. 2.5\% on 7B) at the cost of a 4.2\% decline in average coding benchmark score (where 7B actually improved). We attribute this disparity to MoE-specific training dynamics. In the MoE model's routing scheme, only $\sim$2.4B of 16B parameters are active per token, so each expert sees a small fraction of the training tokens and receives a correspondingly sparser unlearning signal. In a dense model, every parameter participates in every forward pass and thus receives a consistent unlearning signal. In a MoE model, the suppression and reinforcement gradients are fragmented across expert subsets, making consolidation harder and increasing the risk of collateral updates to experts that handle unrelated capabilities. This results in a slight drop in the overall utility of the model compared to other baselines while reducing hallucinations significantly better.

\label{sec:main_results}
\begin{figure*}[!t]
  \centering
  \includegraphics[width=0.9\textwidth]{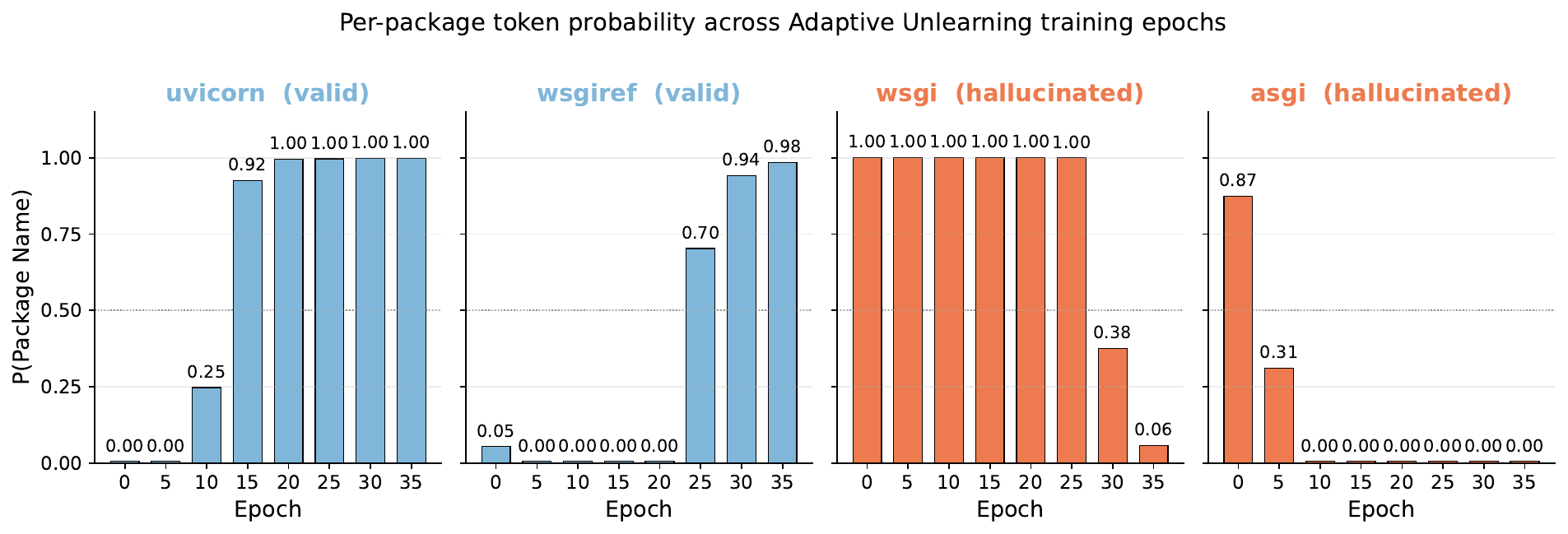}
  \caption{Token-level probabilities during AU for the prompt "How could I write a high-performance Python web framework that supports both WSGI and ASGI, optimized for large-scale API workloads". The model initially recommends two hallucinated packages (orange) that sound plausible for the prompt, but are actually hallucinated. As epochs increase, these packages are suppressed while valid packages (blue) are reinforced. This is a real sample with actual values from our data.}
  \label{fig:probabilities}
\end{figure*}

\section{Ablation Studies}
\label{sec:ablations}

\begin{table*}[h]
\centering
\small
\setlength{\tabcolsep}{4pt}
\begin{tabular}{l cc ccc ccccc}
\toprule
 & \multicolumn{2}{c}{\textbf{Hallucination Rate (\%)} }
 & \multicolumn{3}{c}{$D_{\mathrm{KL}}(\pi_{\theta}\,\|\,\pi_{\text{baseline}})$ }
 & \multicolumn{5}{c}{\textbf{Benchmarks (pass@1)} $\uparrow$} \\
\cmidrule(lr){2-3} \cmidrule(lr){4-6} \cmidrule(lr){7-11}
\textbf{Variant}
 & \textbf{Total$\downarrow$} & \textbf{Abs. Reduction$\uparrow$}
 & \textbf{Code$\downarrow$} & \textbf{Instr.$\downarrow$} & \textbf{Pkg. Avg. $\uparrow$}
& \textbf{HE} & \textbf{HE+} & \textbf{MBPP} & \textbf{MBPP+} & \textbf{Avg.}\\
% \midrule
% \multicolumn{11}{l}{\textit{DeepSeek-1B}} \\
% \midrule
% AU CE-Only         & --           & --          & --               & --               & --               & --             & --     & -- & -- & -- \\
% AU NPO-Only        & --           & --          & --               & --               & --               & --             & --     & -- & -- & -- \\
% AU & --           & --          & --               & --               & --               & --             & --     & -- & -- & -- \\
\midrule
\multicolumn{11}{l}{\textit{DeepSeek-7B}} \\
\midrule
AU CE-Only         & \secondb{12.06} & \secondb{9.154} & \secondb{0.0249} & \best{0.4901} & {0.795} & 70.1 & 62.2 & \best{75.4} & \best{64.8} & 68.12 \\
AU NPO-Only        & 12.42 & 8.81 & \best{0.0203} & \secondb{0.4993} & \secondb{0.8322} & \best{75} & \best{66.5} & 71.7 & 61.6 & \best{68.7}\\
AU                 & \best{2.56} & \best{18.67} & 0.0796 & 0.9905 & \best{1.4211} & \secondb{73.2} & \secondb{64.6} & \secondb{74.3} & \secondb{62.2}  & \secondb{68.57} \\
\midrule
\multicolumn{11}{l}{\textit{DeepSeek-16B}} \\
\midrule
AU CE-Only         & 24.08 & 3.67 & \best{0.0118} & \secondb{0.3959} & 0.0957 & \secondb{68.9} & \secondb{62.8} & 66.7 & 57.4 & \secondb{63.95}  \\
AU NPO-Only        & \secondb{7.11} & \secondb{20.64} & 0.0948 & 0.6263 & \best{1.3153} & 64 & 57.9 & \secondb{68.3} & \secondb{57.7} & 61.97 \\
AU                 & \best{6.13} & \best{21.62} & \secondb{0.0368} & \best{0.299} & \secondb{0.6782} & \best{73.2} & \best{69.5} & \best{81.2} & \best{68.8} & \best{73.17} \\

\bottomrule
\end{tabular}
\vspace{2pt}
\caption{\textbf{Ablation over loss composition.} All three evaluation axes
reported jointly as detailed in \S\ref{sec:ablation_loss}, using the same baseline metrics as Table \ref{tab:main_results}. Best is in bold and second-best is underlined per column.}
\label{tab:ablation_loss_full}
\end{table*}

Having established the overall tradeoff in Table \ref{tab:main_results}, we attribute it to individual components. AU differs from the prior baselines in two ways. It uses a hybrid token-level objective that both reinforces valid package names and suppresses hallucinated ones, and it embeds that objective in an adaptive scaffold of tri-masking, nested epochs, and prompt mutation. We isolate the effect of loss composition first, holding the scaffold fixed, then vary inner epochs and mutation count.

\subsection{Effect of Loss Composition}
\label{sec:ablation_loss}

Figure \ref{fig:probabilities} shows why the hybrid objective works. On the WSGI/ASGI prompt, the hallucinated package names \texttt{wsgi} and \texttt{asgi} start at high probability and fall steadily toward zero, while the valid alternatives \texttt{uvicorn} and \texttt{wsgiref} rise from near zero probability to near certainty. AU is not simply suppressing package mentions indiscriminately. It redistributes probability mass within the package-token space, pushing the model away from hallucinated identifiers and toward valid replacements.

% That qualitative pattern is exactly what a useful hallucination-mitigation mechanism should produce and foreshadows the row-wise tradeoff visible in Table \ref{tab:ablation_loss_full}.

Table \ref{tab:ablation_loss_full} shows that neither loss term alone reproduces AU’s suppression performance. On DeppSeek-7B, removing suppression (CE-Only) leaves hallucination rate at 12.0\%, and removing reinforcement term (NPO-Only) leaves it at 12.42\%. In contrast, the hybrid AU reaches to 2.7\%, a 79\% reduction against both CE-Only and NPO-Only. The two varitations are close to interchangeable in isolation at this scale and far stronger in combination, which is made possible by the tri-mask (\S\ref{subsec:au_loss}).

% These are large gains, and they show that strong suppression emerges only when reinforcement and suppression act together within the same AU scaffold.

% DeepSeek-16B was completely ignored
The scale changes which term carries the result. On DeepSeek-16B, CE-Only reaches only 24.08\% against a base of 27.75\%, while NPO-Only reaches 7.11\%. Suppression does nearly all the work at 16B, and reinforcement alone barely moves the package hallucination rate. The hybrid still improves on both at 6.13\%, while preserving the coding capabilities the best.

% The benchmark block clarifies what each one-sided objective contributes. CE-Only best preserves the MBPP-family benchmarks, reaching 75.4 on MBPP and 64.8 on MBPP+, but fails to meaningfully reduce hallucinations. NPO-Only is strongest on the HumanEval-family benchmarks, scoring 75.0 on HumanEval and 66.5 on HumanEval+, but leaves package hallucination rates above 12\%. Full AU sits between these extremes on utility, 73.2 / 64.6 / 74.3 / 62.2 across HumanEval, HumanEval+, MBPP, and MBPP+, while delivering dramatically stronger hallucination suppression than either isolated variant. The average benchmark scores tell the same story: 68.1 for CE-Only, 68.7 for NPO-Only, and 68.5 for AU. In other words, the hybrid objective does not win by sacrificing utility; it achieves much larger reductions in hallucination rate at essentially the same overall coding-performance level

The benchmark block clarifies what each one-sided objective contributes. CE-Only leads the MBPP family at 75.4 and 64.8, NPO-Only leads the HumanEval family at 75.0 and 66.5, and AU sits between them, with averages of 68.12, 68.70, and 68.57 respectively. AU's hybrid approach buys a fivefold hallucination reduction at essentially no coding cost. On 16B the ordering switches, and AU leads every benchmark column, 73.17 average against 63.95 and 61.97.

The KL results on 7B model show that this suppression is achieved through a larger targeted edit rather than a free improvement. AU induces more drift than either ablation in every domain, code KL rises to 0.079 versus 0.024 and 0.020, instruction KL to 0.990 versus 0.490 and 0.499, and package KL to 1.42 versus 0.795 and 0.8322. The hybrid objective is not the more conservative one. It affects the package distributions the most and achieves the highest suppression. Figure \ref{fig:probabilities} explains why the two signals are complementary. CE-only knows what valid packages to reinforce but lacks a mechanism to actively push fabricated ones down, while NPO-Only suppresses without telling the model what to emit instead.

% \pedram{CE-Only on 16B has package KL of 0.0957, roughly eight times lower than its own 7B value and far below every other row. The row needs rerunning or a footnote, because it is the weakest cell in our ablation and the one a reviewer is most likely to probe.}
\subsection{Effect of Inner Epochs}

\begin{figure}[!t]
  \centering
  \includegraphics[width=0.5\textwidth]{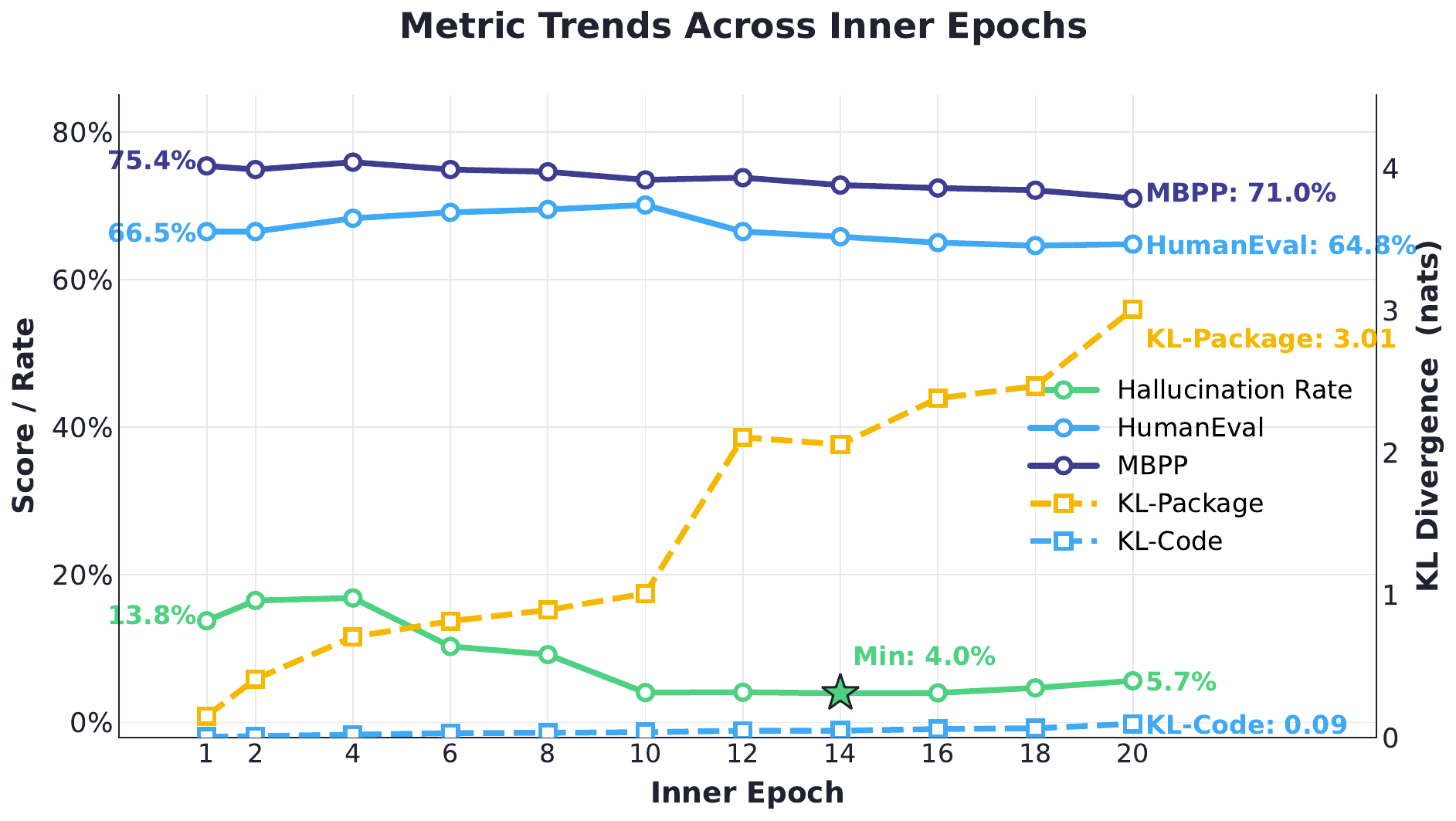}
  \caption{Effect of inner training epochs on hallucination suppression, coding utility, and distributional drift.}
  \label{fig:inner_epoch}
\end{figure}

We vary the number of optimization passes over cached samples before resampling. This ablation tests the claim in \S\ref{subsec:nested} that repeated passes over a fixed batch let the model  consolidate the current suppression signal rather than constantly chasing a moving target.

Figure \ref{fig:inner_epoch} confirms that cached multi-step training is necessary for stable unlearning. With only one inner epoch, hallucination rate remains high at 13.8\%, indicating that resampling too quickly prevents the model from consolidating the suppression signal. Increasing the number of inner epochs improves package hallucination suppression, reaching 4.0\% at 14 inner epochs. From 10 to 20 inner-epochs the hallucination rates vary slightly, increases to 5.7\% at 20 epochs, package KL grows to 3.01, and coding utility softens. These results support the nested-training while identifying the tradeoff between stable collapse and over-specialization.

\subsection{Effect of Prompt Mutations}

\begin{figure*}[!t]
  \centering
  \includegraphics[width=0.75\textwidth]{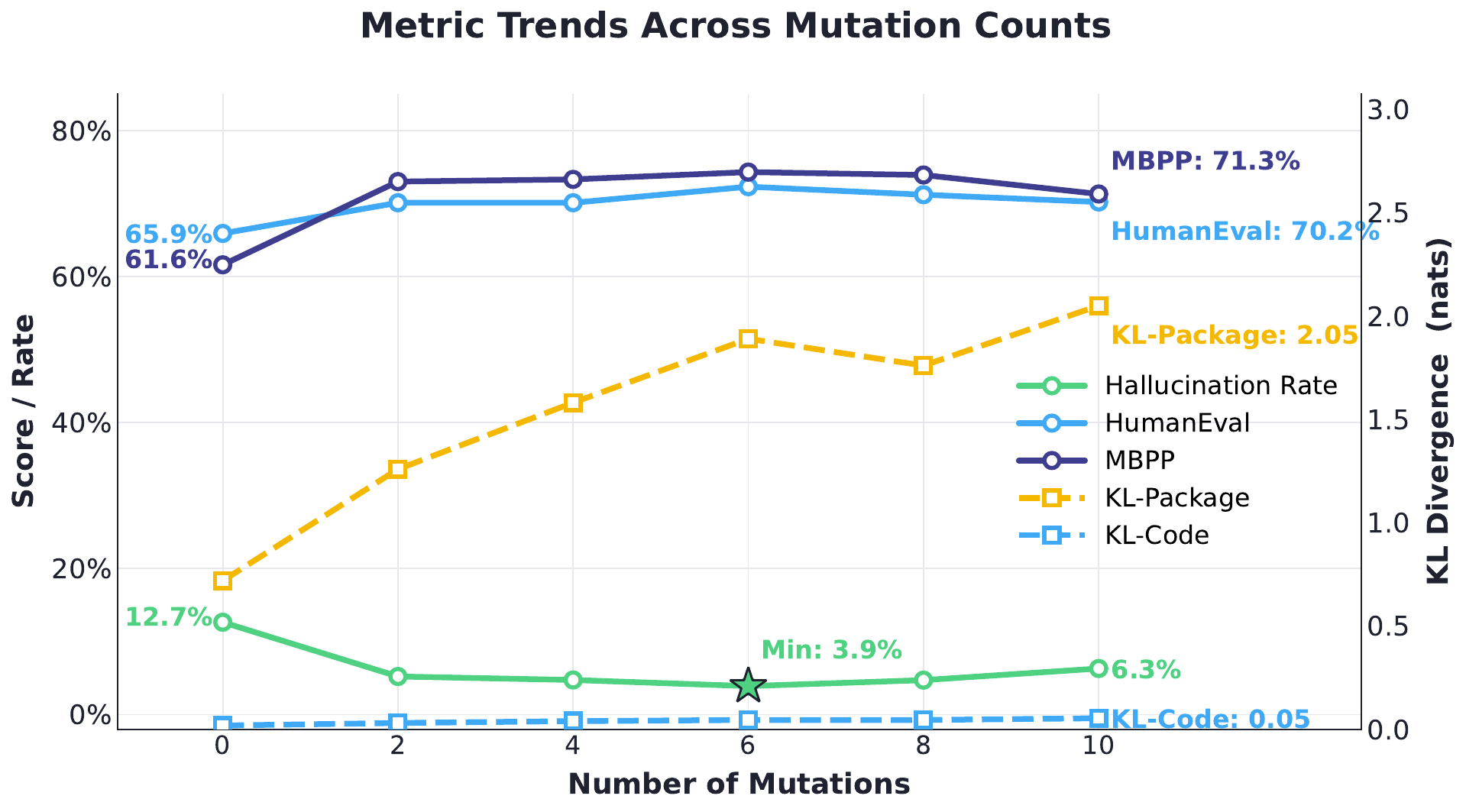}
  \caption{Effect of adaptive prompt mutation count on hallucination rate, coding utility, and distributional drift.}
  \label{fig:mutations}
\end{figure*}

Finally, we vary the number of prompt mutations per run \S\ref{subsec:mutation} to test whether adaptive discovery contributes to generalization beyond the original prompt set. This ablation probes whether AU is learning a reusable suppression pattern or merely overfitting to a static collection of hallucination-inducing prompts.

Figure \ref{fig:mutations} shows prompt mutation is important for both suppression and generalization. Without mutations, hallucination rate remains at 10.7\%. Introducing mutations lowers it quickly, to of 3.9\% at six mutations. Coding utility is preserved or improved across this range, with HumanEval rising from 65.9\% to 70.2\% and MBPP from 61.6\% to 71.3\%. Beyond six mutations the trend reverses. Package hallucination rate rebounds to 6.3\% at ten mutations while package-specific KL continues to rise to 2.05, though code KL remains very small at 0.05. Excessive mutation shifts the package-generation distribution more than necessary.

\section{Limitations and Scope}
\label{sec:limitations}
AU substantially reduces package hallucination at a small coding utility cost. Its assumptions bound where the method applies.

\paragraph{Oracle dependence.} AU requires an oracle labeling generated
text as valid or hallucinated during fine-tuning (not required during standard use). This confines it to domains with deterministic ground truth such as package registries, bibliographic databases, API specifications, and formal systems such as Lean~\cite{lean}. It excludes open-ended factual claims, where automatic verification is itself unsolved. AU also inherits the oracle's blind spot. Our oracle checks that a package name resolves against PyPI, not that the package is used correctly, and packages distributed outside PyPI may be mislabeled. 

\paragraph{Inference-time intervention.}
Crucially, AU does not replace inference-time verification in these domains. It removes the failure from the weights, so \textbf{suppression persists} in deployments where no verifier is in the loop and \textbf{without per-query verification latency}. Filtering packages at inference time leaves the underlying rate unchanged and must run on every generation, which requires registry access at inference time and adds per-query latency. Regeneration after rejection can hallucinate again.

\paragraph{Oracle poisoning.} In the slopsquatting setting, an attacker who pre-registers hallucinated names inverts the oracle's signal. Those names move from the forget partition to the retain partition, and AU reinforces what it was deployed to suppress. The failure is silent, since nothing in the loop distinguishes an adversarial registration from a legitimate one.
Training data must therefore be drawn from a registry snapshot audited for such injections.

\paragraph{Token-disjointness assumption.} Tri-masking presumes each token belongs to exactly one of the valid, hallucinated, and context populations. Package identifiers make this clean since they are lexically isolated and rarely overlap surrounding code. The assumption weakens when a hallucinated package shares tokens with valid content. A suppressed package name may collide with a variable or symbol in an existing codebase, or a multi-token fabrication embedded in an otherwise correct sentence. We evaluate on standalone generation tasks and do not measure this collision case. Extending tri-masking to overlapping or soft-assigned spans remain an open work for the future.

\paragraph{Evaluated scope.} We evaluate two DeepSeek-Coder models, 7B and 16B on Python coding applications. Hallucination behavior varies across model families and scales, and our results should not be read as characterizing LLMs in general. Full-parameter fine-tuning at frontier scale exceeded our compute budget. Extensions to other ecosystems (npm, vcpkg, Conan), model families, and general-purpose assistants remain future work.

\section{Conclusion}
\label{sec:conclusion}
Package hallucinations in code-generating LLMs are an exploitable supply-chain attack surface: each fabricated identifier is a name an adversary can register and weaponize against developers and autonomous agents that trust the model's output. We presented \textbf{Adaptive Unlearning (AU)}, a post-deployment framework that frames this problem as machine unlearning over an open-ended forget set and addresses it through a token-level hybrid objective—reinforcing valid generations and suppressing hallucinated ones via a tri-mask partition—wrapped in an adaptive prompt-mutation loop that continuously surfaces new hallucination-inducing contexts.

AU reduces total package hallucination rates by 88\% on DeepSeek-7B and 78\% on DeepSeek-16B while preserving coding utility on established coding benchmarks. KL-divergence analysis confirms the induced change is concentrated in the package-generation distribution and largely absent on ordinary code generation. AU outperforms gradient ascent, NPO, and PMC jointly across suppression, utility, and targeted-drift axes—the combination a deployable post-deployment defense requires.

Ablations identify the load-bearing design choices: neither the reinforcement nor the suppression term suffices alone, with one-sided variants leaving hallucination rates above 12\%; the nested-epoch structure has a clear sweet spot for sample-cache reuse; and prompt mutation contributes measurable generalization beyond the original prompt set.

AU offers a practical path for vendors to address user-reported hallucinations between release cycles, without full retraining or inference-time verification overhead. We release all artifacts (\S\ref{sec:open_science}) and identify cross-language generalization, semantic-correctness oracles, and applying AU to general-purpose LLMs as the most consequential extensions.

% \claude{Writing pass requested by Reviewer A across all sections: break up
% long sentences (their two quoted examples are from the PMC discussion and
% the Tan et al.\ description), cut adjectives like ``critical innovation''
% and ``surgical,'' and remove informal phrasings such as the quoted
% ``don't hallucinate for these exact prompts'' sentence.}

%-------------------------------------------------------------------------------
\section{Acknowledgments}
We disclose that generative AI tools (specifically, Anthropic's Claude) were used in the preparation of this work in two limited capacities: (i) as a writing aid for paraphrasing, copy-editing, and improving the clarity of author-drafted text, and (ii) as a coding assistant for debugging components of the experimental and evaluation pipelines. The AI tool was not used to generate the research ideas, design the methodology, conduct the experiments, analyze the results, or produce any of the scientific claims advanced in this paper. All technical content, experimental design, results, and conclusions are the work of the authors, who take full responsibility for the contents of the paper.
%-------------------------------------------------------------------------------
% USENIX uses an unnumbered section rather than ACM's \begin{acks} environment.
% Leave empty/commented for anonymous submission.
% This paper was edited for grammar using [Tool Name].

%-------------------------------------------------------------------------------
\paragraph{Access during double-blind review.} All released artifacts are available at an anonymized repository: \url{https://anonymous.4open.science/r/Adaptive-Unlearning-952E}. The repository requires no credentials and contains no identifying information about the authors or affiliated institutions. Reviewers can clone it directly. Upon acceptance, the artifacts will be migrated to a permanent, non-anonymous repository under a permissive open-source license, and the canonical URL will be added to the camera-ready version.
%-------------------------------------------------------------------------------
%-------------------------------------------------------------------------------
\bibliographystyle{plain}

\bibliography{ref}

%% Appendices
\appendix
\section{Open Science}
%-------------------------------------------------------------------------------
\label{sec:open_science}

In accordance with the USENIX Open Science policy, we release the artifacts required to reproduce and evaluate the core contributions of this work.

\paragraph{Artifacts released.} We release the following artifacts:
\begin{itemize}
    \item \textbf{Source code} for the Adaptive Unlearning training framework, including the implementations of all baselines (Gradient Ascent, NPO, PMC) and ablated variants (AU CE-Only, AU NPO-Only, Adaptive Unlearning) used in our experiments.
    \item \textbf{Tri-mask construction and registry-oracle code} implementing the hallucination-detection pipeline against the PyPI snapshot.
    \item \textbf{The curated prompt set} of 40 code-generation prompts used for training and evaluation, including the full set of mutated variants generated during the adaptive loop.
    \item \textbf{Evaluation harnesses} for the hallucination-rate, KL-drift, and EvalPlus utility benchmarks, including all preprocessing and aggregation scripts.
    \item \textbf{Configuration files and hyperparameter sweeps} for every reported experimental run, sufficient to reproduce each table and figure in the paper.
    \item \textbf{Documentation} including a README with environment setup instructions, expected runtimes, and per-experiment reproduction commands.
\end{itemize}

% \paragraph{Access during double-blind review.} All released artifacts are available at an anonymized repository: \url{https://anonymous.4open.science/r/Adaptive-Unlearning-952E}. The repository requires no credentials and contains no identifying information about the authors or affiliated institutions. Reviewers can clone it directly. Upon acceptance, the artifacts will be migrated to a permanent, non-anonymous repository under a permissive open-source license, and the canonical URL will be added to the camera-ready version.

% \paragraph{Sufficiency for evaluation.} The released code, prompts, configurations, and evaluation harnesses are sufficient for reviewers to independently reproduce the central claims of this paper, including the over 80\% reduction in package hallucination rates and the preservation of utility on standard coding benchmarks under the EvalPlus framework.

\section{Ethical Considerations}
\label{sec:ethics}

This work raises no ethical concerns requiring institutional review or special handling. The research does not involve human subjects, personally identifiable information, or sensitive user data. All experiments are conducted on publicly available, openly licensed code-generation models (DeepSeek-Coder-7B-Instruct-v1.5 and DeepSeek-Coder-V2-Lite-Instruct) using model-generated synthetic data; no human-annotated corpora were collected or used. The threat model we address---slopsquatting via package hallucinations---is already well-documented in the published literature~\cite{spracklen2025}, and our work develops a defense rather than a new attack, with the explicit goal of reducing the security exposure of LLM-assisted software development. We have not introduced any novel attack technique that adversaries could repurpose, and our released artifacts (Section~\ref{sec:open_science}) consist entirely of defensive tooling. Accordingly, no responsible-disclosure process or coordination with affected vendors was required for this work.
\section{Prompt Example}
\label{app:app_prompts}
Figure~\ref{fig:example_prompt} shows a representative prompt from our evaluation set.

\begin{figure}[t]
\centering
\fcolorbox{black}{gray!15}{%
\parbox{0.95\linewidth}{%
\small\color{black!85}
Generate Python code that composes multiple
large language models (LLMs) to build an
application that can perform a variety of
tasks, such as natural language understanding,
image recognition, and decision making,
by using modular architectures and interfaces.
}}
\caption{Representative code-generation prompt from the evaluation set.}
\label{fig:example_prompt}
\end{figure}
\section{Hyperparameter Details}
\label{app:hyperparams}
\paragraph{Hyperparameters.} Unless otherwise stated, we fix the following configuration for Adaptive Unlearning:
\begin{itemize}
    \item Batch size: 8
    \item Maximum sequence length: 1024 tokens
    \item Mutations per prompt: 6
    \item Exhaustion threshold: 5 outer epochs or 0 hallucinations
    \item Samples generated and cached per outer epoch: 5
\end{itemize}
Across our hyperparameter exploration, we evaluated several promising loss-weight configurations:
\begin{center}
\small
\begin{tabular}{cccl}
\toprule
$\lambda_{\text{retain}}$ & $\lambda_{\text{forget}}$ & $\lambda_{\text{reg}}$ & \\
\midrule
1.25 & 1.00 & 1.00 & \\
1.20 & 0.50 & 1.00 & \\
1.25 & 0.75 & 1.00 & \\
1.00 & 1.25 & 1.00 & (default) \\
\bottomrule
\end{tabular}
\end{center}
The latter two configurations consistently outperformed the alternatives, with $(1.00, 1.25, 1.00)$ producing slightly better Adaptive Unlearning results across repeated runs. Per-run learning rates, outer/inner-epoch counts, and the specific loss-weight settings used for each baseline and ablation are reported in Appendix~\ref{app:hyperparams} (Table~\ref{tab:hyperparams_full}).

Table~\ref{tab:hyperparams_full} reports the exact hyperparameter configuration for every experimental run reported in Tables~\ref{tab:main_results} and~\ref{tab:ablation_loss_full}. All runs use full-parameter fine-tuning (FP) on AMD MI200 / MI300 GPUs. Where the planned outer-epoch budget was not reached because the run hit the 24/12-hour wall-clock limit, both the planned and used checkpoints are mentioned. For the reported AU on DeepSeek-16B the best reported result is a restart from a similar configured run at checkpoint 60 for another 12 hours. 

\begin{table*}[t]
\centering
\small
\setlength{\tabcolsep}{4pt}
\renewcommand{\arraystretch}{1.15}
\begin{tabular}{l l c r r c c c c}
\toprule
\textbf{Model} & \textbf{Method} & \textbf{LR} & \textbf{Outer} & \textbf{Inner} & \textbf{$\lambda_{\text{retain}}$} & \textbf{$\lambda_{\text{forget}}$} & \textbf{$\lambda_{\text{reg}}$} & \textbf{Checkpoint Num.}\\
\midrule
\multicolumn{9}{l}{\textit{deepseek-coder-7b-instruct-v1.5}} \\
\midrule
DeepSeek-7B  & GA                 & $1\!\times\!10^{-5}$ & 1     & 10 & --   & -- & -- & final    \\
% DeepSeek-7B  & NPO                & $1\!\times\!10^{-5}$ & 1     & 1  & --   & -- & -- & final     \\
DeepSeek-7B  & NPO                & $1\!\times\!10^{-5}$ & 1     & 60 & --   & -- & -- & final     \\
DeepSeek-7B  & PMC                & $1\!\times\!10^{-5}$ & -- & 10 & --   & -- & -- & 25     \\
DeepSeek-7B  & AU-CE-Only       & $1\!\times\!10^{-5}$ & 10 & 5  & --   & -- & -- & 15  \\
DeepSeek-7B  & AU-NPO-Only       & $1\!\times\!10^{-5}$ & 10 & 10 & --   & -- & -- & 15  \\
DeepSeek-7B  & AU & $5\!\times\!10^{-5}$ & 10 & 20 & 1.00 & 1.25 & 1.00 & 60 \\
\midrule
\multicolumn{9}{l}{\textit{DeepSeek-Coder-V2-Lite-Instruct}} \\
\midrule
DeepSeek-16B & GA                 & $1\!\times\!10^{-5}$ & 1      & 10    & --   & -- & -- & final\\
DeepSeek-16B & NPO                & $1\!\times\!10^{-5}$ & 1      & 60    & --   & -- & -- & final\\
DeepSeek-16B & PMC                & $2\!\times\!10^{-5}$ & --     & 20 & --   & -- & -- & 20 \\
DeepSeek-16B & AU-CE-Only        & $1\!\times\!10^{-5}$ & 10     & 10 & --   & -- & -- & 30 \\
DeepSeek-16B & AU-NPO-Only       & $5\!\times\!10^{-5}$ & 10     & 20 & --   & -- & -- & 60 \\
DeepSeek-16B & AU & $2\!\times\!10^{-5}$ & 10     & 20 & 1.00 & 1.25 & 1.00 & 50 restarted\\
\bottomrule
\end{tabular}
\vspace{2pt}
\caption{\textbf{Per-run hyperparameter configurations.} LR = learning rate;
Outer / Inner = number of outer and inner epochs (notation {$N_{\text{planned}}$} when the run hit the 24-hour wall-clock limit for 7B models and 12-hour wall-clock limit for 16B models the checkpoint that yielded best results is mentioned).
$\lambda_{\text{retain}}$, $\lambda_{\text{forget}}$, $\lambda_{\text{reg}}$ are the loss-component weights from \S\ref{subsec:au_loss}; -- indicates the term is not used by that method. All runs use full-parameter fine-tuning. Method names follow \S\ref{sec:baselines}.}
\label{tab:hyperparams_full}
\end{table*}

\section{Partial Model Collapse}
\label{app:pmc_formula}
PMC is a self-training unlearning mechanism that induces targeted distribution collapse by repeatedly fine-tuning a model on its own generated outputs, rather than on explicit unlearning targets. We summarize the core formulation here and refer to prior work for full
theoretical analysis~\cite{scholten2025pmc}.

Let $\pi_\theta(y \mid p)$ denote the model distribution over outputs $y$
conditioned on a prompt $p$.
We assume a retain prompt distribution $p_r(p)$ and a forget prompt
distribution $p_f(p)$ with disjoint support.
PMC defines an iterative update of the model distribution via:
\begin{equation}
\begin{split}
\pi_{\theta_{t+1}}
= \arg\max_{\pi} \Bigg[
&\lambda \, \mathbb{E}_{(p,y)\sim p_r}
\big[\log \pi(y \mid p)\big] \\
&+ \mathbb{E}_{p\sim p_f}
\mathbb{E}_{\hat{y} \sim \mathcal{C}(\pi_{\theta_t},p)}
\big[\log \pi(\hat{y} \mid p)\big]
\Bigg]
\end{split}
\label{eq:pmc_objective}
\end{equation}
where $\lambda$ balances utility preservation and unlearning, and
$\mathcal{C}(\pi_{\theta_t},p)$ denotes a curation process that samples
multiple candidate outputs from the current model $\pi_{\theta_t}$ and
selects one according to a scoring function.

Intuitively, the retain term preserves general model utility, while the
forget term trains the model on its own preferred generations for forget
prompts. 

Repeated optimization of~\eqref{eq:pmc_objective} causes the
conditional output distribution $\pi_{\theta}(y \mid p)$ to collapse
toward a restricted subset of responses for forget prompts, effectively
removing undesired information without explicitly optimizing against
ground-truth unlearning targets~\cite{scholten2025pmc}.

Prior work shows that, in the absence of approximation error, this
iterative process converges such that the expected reward on forget prompts approaches 
its maximum and variance vanishes, while the model distribution on 
retain prompts converges to the retain distribution.

\section{Initial Coding Prompts}
\label{app:app_initial_prompts}

These are the coding prompts that our model begins training on. These prompts are then randomly mutated during training to produce similar, but distinct, questions. 

\begin{enumerate}
    \item ``Generate Python code that imports the required libraries and creates a composite Transformer model using the XFormers library, allowing for flexible and efficient construction of complex Transformer architectures."
    \item ``Generate Python code that implements a generative adversarial network (GAN) for face restoration, using convolutional neural networks (CNNs) to generate high-quality images from low-resolution input faces."
    \item ``Generate Python code that implements a signal system for Django queries, allowing developers to intercept and modify querysets before they are executed, including hooks for pre-sql injection, post-sql injection, and custom queryset modification."
    \item ``Generate Python code that implements an efficient object detection model using PyTorch, featuring anchor boxes, bounding box regression, and class probabilities, with optional augmentations and refined anchors for improved accuracy."
    \item ``Generate Python code that uses scikit-learn's GridSearchCV to perform beta calibration on a given dataset, using a specified model and hyperparameter grid."
    \item ``Generate Python code that composes multiple large language models (LLMs) to build an application that can perform a variety of tasks, such as natural language understanding, image recognition, and decision making, by using modular architectures and interfaces."
    \item ``Generate Python code that parses a Python script and generates a full syntax tree, allowing for easy refactoring and modification of the code."
    \item ``Generate Python code that wraps systemd interfaces, providing a simple and consistent API for interacting with systemd services, units, and snapshots."
    \item ``Generate Python code that imports the necessary modules and defines a custom Slack provider class with the required methods to send messages and attachments to a Slack channel, similar to the apache-airflow-backport-providers-slack package."
    \item ``Generate Python code that uses Aspose.Words to read, edit, and save Word documents, Excel spreadsheets, and PDF files without requiring Microsoft Office or any other external dependencies."
    \item ``Generate Python code that uses TensorFlow and Keras to build and train a deep neural network for image classification, utilizing transfer learning and pre-trained models to achieve high accuracy."
    \item ``Generate Python code that imports the required modules and annotates the types for the MediaStoreData service in boto3"
    \item ``Generate Python code that imports the necessary modules and defines a class with common attributes for interacting with Amazon DynamoDB, similar to the `pynamodb` package."
    \item ``Generate Python code that imports the necessary libraries and utilizes System V IPC primitives (semaphores, shared memory, and message queues) to facilitate inter-process communication and synchronization."
    \item ``Generate Python code that uses the `pandas` and `scipy` libraries to perform genomic sequence analysis, including fasta file parsing, genome assembly, and variant calling."
    \item ``Generate Python code that creates a game window with a grid of buttons, handles user input to move a character around the grid, and implements collision detection to prevent the character from moving outside the grid or into other objects."
    \item ``Generate Python code that implements an ultra-reliable, fast ASGI+WSGI framework for building data plane APIs at scale, utilizing the latest performance optimizations and scalability features of the Python ecosystem."
    \item ``Generate Python code that installs the necessary packages for PyQt5 to function properly on a system, including the Qt development libraries and dependencies."
    \item ``Generate Python code that imports the necessary modules and creates a CRF suite wrapper similar to scikit-learn, allowing for efficient and accurate statistical modeling and prediction."
    \item ``Generate Python code that uses the TensorFlow and Apache Airflow libraries to create a workflow for training, validating, and deploying machine learning models."
\end{enumerate}

\begin{figure*}[!t]
  \centering
  \includegraphics[width=0.95\textwidth]{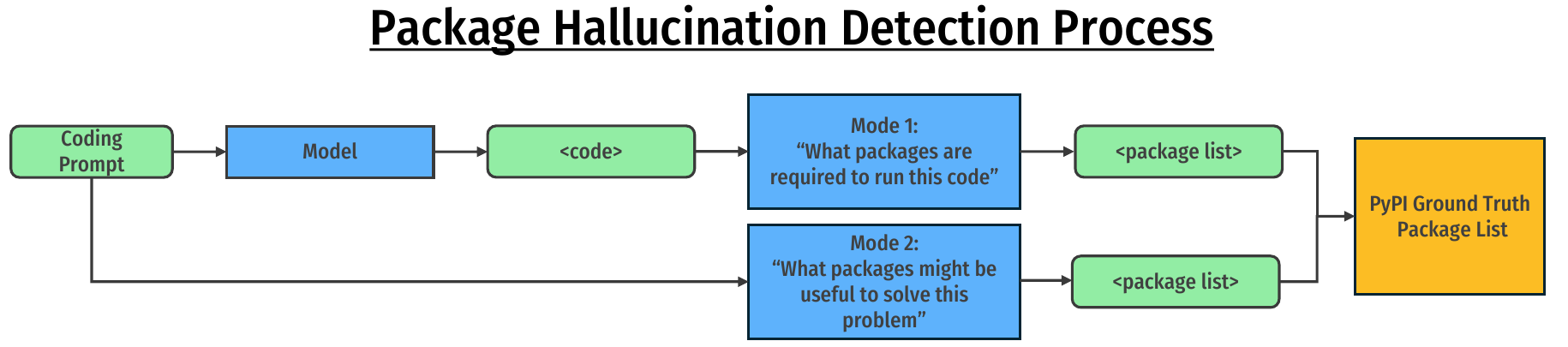}
  \caption{Visual depiction of the package detection pipeline to test for package hallucinations, as described in \S\ref{sec:domain}: Each coding prompt is passed to the model to produce code, then dispatched to two distinct package-elicitation queries: \emph{Mode 1} asks for the packages required to run the generated code, and \emph{Mode 2} asks for packages that would be useful in solving the original task. Each resulting package list is parsed, de-duplicated, and resolved against a snapshot of the PyPI registry; any identifier that fails to resolve is labeled as a hallucination and routed into the tri-mask construction (\S\ref{sec:tri-masking}).}
  
  \label{fig:package_detection}
\end{figure*}

\section{Hallucination Detection Pipeline}
\label{app:app_detection}
Figure~\ref{fig:package_detection} illustrates the dual-mode package elicitation procedure that underlies both AU's training-time tri-mask construction and the test-time hallucination-rate evaluation. The two modes capture complementary failure surfaces: Mode~1 stresses runtime correctness (a hallucinated import is a deterministic install-time failure and an immediate slopsquatting target), while Mode~2 stresses recommendation quality (where hallucinated package names appear most frequently and are the hardest for users to vet, since ``useful'' suggestions are not validated against an actual import graph). Reporting hallucination rates separately for the two modes, as we do in our main results, prevents either elicitation regime from dominating the headline number and exposes methods that suppress one type of hallucination while leaving the other intact.

The PyPI ground-truth list is constructed from a static snapshot of the registry collected prior to all experiments and held fixed across training and evaluation. Edge cases of the registry-based oracle, most notably, legitimate packages distributed outside PyPI, are discussed in \S\ref{sec:limitations}.

\end{document}